\documentclass{emulateapj}
\usepackage{apjfonts}

\renewcommand{\vec}[1]{\mbox{\boldmath $\displaystyle #1$}}
\newcommand{\grad}{\vec{\nabla}}

\newcommand{\vdot}{\vec{\cdot}}
\newcommand{\vcross}{\vec{\times}}
\newcommand{\divr}{\grad\vdot\,}
\newcommand{\curl}{\grad\vcross\,}
\newcommand{\be}{\begin{eqnarray}}
\newcommand{\ee}{\end{eqnarray}}

\begin{document}

\title{Magnetic Field Evolution in Neutron Star Crusts Due to the Hall Effect
and Ohmic Decay}

\author{Andrew Cumming} \affil{Hubble Fellow, UCO/Lick Observatory and
Department of Astronomy and Astrophysics, University of California,
Santa Cruz, CA 95064}
\author{Phil Arras} \affil{NSF AAPF Fellow, Kavli Institute for
Theoretical Physics, Kohn Hall, University of California, Santa
Barbara, CA 93106}
\author{Ellen Zweibel} \affil{ Departments of Astronomy and Physics and
Center for Magnetic Self-Organization, University of
Wisconsin, 475 North Charter Street, Madison, WI 53706}

\begin{abstract}
We present calculations of magnetic field evolution by the Hall effect
and Ohmic decay in the crust of neutron stars (NSs). In accreting NSs,
Ohmic decay is always the dominant effect due to the large resistivity. In
isolated NSs with relatively pure crusts, the Hall effect dominates Ohmic decay after a time $t_{\rm switch}\simeq 10^4\ \mathrm{yr}\ B_{12}^{-3}$, where $B_{12}$ is the magnetic field strength in units of $10^{12}\ \mathrm{G}$. We compute the evolution of an initial field distribution by Ohmic decay, and give approximate analytic formulas for both the surface and interior fields as a function of time. Due to the strong dependence of $t_{\rm switch}$ on $B_{12}$, early Ohmic decay can
alter the currents down to the base of the crust for $B\sim 10^{11} \mathrm{G}$, neutron drip for $B\sim 10^{12}\ \mathrm{G}$, and near the top of the crust for $B\gtrsim 10^{13}\ \mathrm{G}$, respectively.  We then discuss magnetic field evolution by the Hall effect. Several examples are given to illustrate how an initial field configuration evolves. Hall wave eigenfunctions are computed,
including the effect of the large density change across the crust.
We estimate the response of the crust to the magnetic stresses induced by
Hall waves, and give a detailed discussion of the boundary conditions at the solid-liquid interface. Finally, we discuss the implications for the Hall cascade proposed by Goldreich \& Reisenegger.
\end{abstract}



\section{Introduction}

Our picture of neutron star magnetic field evolution comes from studies of radio pulsars, accreting neutron stars, and most recently, magnetars. The known radio pulsars fall into two general classes: young ($\lesssim 10^7\ \mathrm{yrs}$) pulsars with spin periods $\sim 1\ \mathrm{s}$ and $\sim 10^{12}\ \mathrm{G}$ magnetic fields, and the millisecond radio pulsars, which have inferred magnetic fields as low as $10^8\ {\rm G}$. Whereas most radio pulsars are isolated objects, the millisecond pulsars are predominantly in binaries and have undergone a period of accretion. This has led to the suggestion that millisecond pulsars form by accretion, which both spins up the neutron start to short rotation periods, and also acts to reduce the magnetic field (see Bhattacharya 1995 for a review). 

This ``recycling'' scenario is consistent with observations of accreting neutron stars. Young accreting neutron stars in high mass X-ray binaries (HMXBs) are observed as X-ray pulsars with inferred $\sim 10^{12}\ {\rm G}$ magnetic fields, confirmed by detection of cyclotron lines in their X-ray spectra (Tr\"umper et al.~1978). Old accreting neutron stars in low mass X-ray binaries (LMXBs) are generally not X-ray pulsars, implying no magnetic disruption of the accretion disk in those systems, and magnetic fields $\lesssim 10^9$--$10^{10}\ {\rm G}$ (see Cumming, Zweibel, \& Bildsten 2001 for a recent discussion). Recently, 5 millisecond accreting X-ray pulsars have been discovered in transient LMXBs, with spin periods between $180$ and $410\ \mathrm{Hz}$, and inferred magnetic fields $10^8$---$10^9\ \mathrm{G}$, exactly in the range expected for millisecond pulsar progenitors (Chakrabarty et al.~2003).

An additional possibility is that neutron star magnetic fields decay with age. Field evolution in isolated radio pulsars has been studied using their space and velocity distributions, as well as the statistical distribution in the $P$-$\dot P$ diagram.  This probes field evolution on the timescale of the ages of radio pulsars ($\lesssim 10^7\ {\rm yrs}$). At first, it was argued that exponential field decay on a few million year timescale was required
to understand the $P$--$\dot P$ distribution (Ostriker \& Gunn 1969; Gunn
\& Ostriker 1970; Narayan \& Ostriker 1990), but the work of Bhattacharya
et al.~(1992) and others in the 90's argued that in fact no field decay occurs over the lifetime of a radio pulsar. However, this issue remains somewhat open: recently, several authors have again raised the suggestion that neutron star magnetic fields evolve on a timescale something like 10 million years (Cordes \& Chernoff 1998;
Tauris \& Manchester 1998; Tauris \& Konar 2001; Gonthier et al.~2002).

The discovery of magnetars has complicated this picture, but also provided a new opportunity to study magnetic field evolution on relatively short timescales. Observed as the young $\lesssim 10^4\ \mathrm{yr}$ soft gamma repeaters (SGRs) and anomalous X-ray pulsars (AXPs), these sources are believed to be directly powered by decay of ultrastrong $\sim 10^{14}$--$10^{15}\ \mathrm{G}$ magnetic fields. Long term release of magnetic energy may act to delay the cooling of the neutron star (Thompson \& Duncan 1996; Heyl \& Kulkarni 1998; Colpi, Geppert, \& Page 2000; Arras, Cumming, and Thompson 2004), while rapid release of magnetic stresses that build in the neutron star crust (Duncan \& Thompson 1992; Thompson \& Duncan 1995, 1996) may power the frequent gamma-ray bursts observed from SGRs, and now also AXP 1E 1048.1-5937 (Gavril, Kaspi, \& Woods 2002).

These ideas and observations have motivated a large range of theoretical work on neutron star magnetic field evolution (e.g.~see Bhattacharya \& Srinivasan 1995 for a review).  In this paper, we discuss the evolution of currents in the neutron star crust, which evolve due to Ohmic decay, and by the Hall effect. The importance of currents in the crust is that they decay much more quickly than currents in the core. They therefore play an important role in models of accretion-induced field evolution: either the currents are assumed to flow only within in the crust, for example produced after the birth of the neutron star by thermomagnetic effects (Blandford, Applegate, \& Hernquist 1983), or magnetic flux may be expelled from the core into the crust due to vortex-fluxoid interactions (e.g.~Srinivasan et al.~1990; Ruderman, Zhu, \& Chen 1998). At higher field strengths, Hall effects most likely are responsible for building up stresses in magnetar crusts which then are released during SGR and AXP bursts.

Previous investigations into Ohmic decay have focused mainly on the decay of the surface field, and involved calculating Ohmic decay eigenmodes for the crust (Sang \& Chanmugam 1987), self-similar solutions near the surface of the star (Urpin, Chanmugam, \& Sang 1994), or direct time-integrations of the Ohmic diffusion equation for an initial distribution of currents (Urpin \& Muslimov 1992; Urpin \& Konenkov 1997; Page, Geppert, \& Zannias 2000). For accreting stars, the evolution due to Ohmic decay for fields supported by crustal currents has been calculated in great detail (e.g., Geppert \& Urpin 1994; Urpin, Geppert, \& Konenkov 1998; Konar \& Bhattacharya 1997). We revisit Ohmic decay in this paper, calculating both Ohmic decay modes and self-similar solutions, and show that they are related in a simple way. As well as constraining the ``initial condition'' for the Hall effect, this allows a general way to understand the distribution of currents in the crust at a given age, as well as the decline of the surface field.

The evolution by the Hall effect is much less well-understood, being non-linear and therefore much more complex. Although the Hall effect itself is non-dissipative, Goldreich \& Reisenegger (1992, hereafter GR) proposed that it leads to dissipation through a turbulent ``Hall cascade'', magnetic energy cascading from large to small scales where it dissipates by Ohmic decay in the Hall time associated with the largest lengthscale. Recent studies by Biskamp and collaborators (Biskamp et al.~1999) of Whistler turbulence indeed show a clear cascade of energy over more than an order of magnitude in lengthscale. Other authors have also demonstrated transfer of magnetic energy between different scales (Muslimov 1994; Naito \& Kojima 1994; Shalybkov \& Urpin 1997; Urpin \& Shalybkov 1999; Hollerbach \& R\"udiger 2002). In a different approach, Vainshtein, Chitre, \& Olinto (2000) included the effect of the density gradient in a simple model. Most recently, Rheinhardt \& Geppert (2000), Geppert \& Rheinhardt (2002), and Rheinhardt, Konenkov, \& Geppert (2003) discuss a ``Hall drift instability'' which can lead to non-local transfer of energy to small scales.

In this paper, we study the implications of realistic crust models for evolution of crustal currents. In \S 2, we calculate the timescales associated with Ohmic decay and the Hall effect for realistic crust models. We discuss the relative importance of Ohmic decay and the Hall effect at different stages of a neutron star's life. In \S 3, we consider Ohmic decay in detail. We show that previous studies of Ohmic decay can be understood in a simple way. In \S 4, we discuss the evolution due to the Hall effect. We start with a simple discussion of the initial evolution of a dipole field which gives a simple physical picture. We then calculate the properties of linear Hall waves in a realistic crust model. We clarify the physical nature of the ``Hall drift instability'' and comment on its relevance for the overall field evolution. We summarize our main results in \S 5. In the Appendix, we discuss the behavior of Hall waves at the boundaries of the crust.


\section{Timescales for the Hall Effect and Ohmic Decay}

In this section, we estimate the timescales of evolution by the Hall
effect and Ohmic decay in both isolated and accreting neutron stars,
and discuss the regimes where the Hall effect is important. Both the
Hall and Ohmic timescales are most sensitive to the lengthscale $L$
over which the magnetic field and crust properties vary. In this
section we make a simple estimate by setting $L$ equal to the local
pressure scale height $H$. This allows a comparison between the two
processes, and sets the scene for the more detailed discussion in \S \ref{sec:ohmicevolution} and
\ref{sec:fieldevolution}.

\subsection{Calculation of the Hall timescale}
\label{sec:halltime}

The evolution of the magnetic field is given by Faraday's law,
\begin{equation}\label{eq:faraday}
{\partial\vec{B}\over\partial t}=-c\curl\vec{E},
\end{equation}
together with Ohm's law
\begin{equation}\label{eq:ohm}
\vec{E}=- \frac{1}{c} \vec{v} \times \vec{B} +
{\vec{J}\over\sigma}+{\vec{J}\vcross\vec{B}\over n_e e c},
\end{equation}
where $\sigma$ is the electrical conductivity, and the current density
is $\vec{J}=(c/4\pi)(\curl\vec{B})$ (see GR for a derivation). The
first term in equation (\ref{eq:ohm}) describes advection of the field
by the fluid, the second term describes Ohmic dissipation, and the
third term is the Hall electric field\footnote{We do not discuss the
battery or thermoelectric terms in this paper (e.g.~Blandford et al.~1983) since they are suppressed by a factor
$(k_BT/E_{F,e})^2$ for a degenerate gas, and are important in the crust only for
very young neutron stars.}. Neglecting the advection and Ohmic terms,
the magnetic field evolution is given by
\begin{equation}\label{eq:inducthall}
\frac{ \partial \vec{B}}{\partial t} = 
-\curl \left( \frac{\vec{J}\vcross\vec{B}}{n_e e} \right),
\label{eq:hall}
\end{equation}
with a timescale
\begin{equation}\label{eq:thall}
t_{\rm Hall}={n_eeL\over J}={4\pi n_eeL^2\over cB},
\end{equation}
where $L$ is the typical lengthscale over which $B$, $J$, and $n_e$
vary.

\begin{figure}
\epsscale{1.3}\plotone{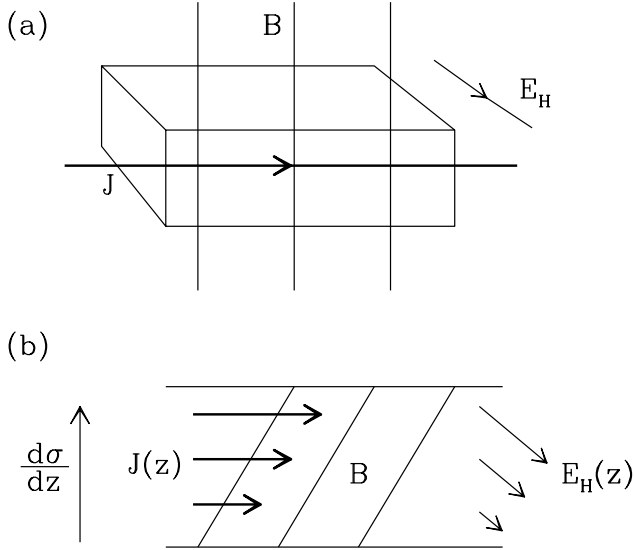}
\caption{Simple example of the Hall effect: (a) the usual laboratory
set up in which the Hall electric field balances the magnetic force on
the conducting electrons; (b) if the conductivity varies with height,
the electron velocity depends on height, shearing the magnetic field
on a timescale $t_{\rm Hall}=L/v_e=n_eeL/J$.\label{fig:shear}}
\end{figure}

The basic physics of equation (\ref{eq:inducthall}) is that the magnetic
field is frozen to the electrons. In ideal MHD, the electron and ion 
velocities are the same - both follow the $\vec{E}\times\vec{B}$ drift. In
Hall MHD, the ions follow separate dynamics. In a plasma or liquid, this
only happens on scales of order the ion skindepth $\delta_i\equiv v_A/\omega_{ci}$ (or, in a high $\beta$ plasma, scales of order the ion gyroradius
$v_i/\omega_{ci}$). In solids, the ions are immobilized by lattice forces
and Hall effects are important on much larger scales.
This is the case in the neutron star crust, which elastically adjusts as the 
field evolves, balancing the $\vec{J}\vcross\vec{B}$ forces with shear 
stresses. We discuss the issue of force balance in more detail in \S 4, 
where we consider the behavior of wave-like solutions at the boundary between the solid crust and fluid core/ocean. This is
illustrated by Figure \ref{fig:shear} which shows a variant of the
usual laboratory demonstration of the Hall effect. A current-carrying
conductor is placed in an external magnetic field, with the current
perpendicular to the field.  The conducting electrons, which move with
velocity $\vec{v_e}=-\vec{J}/n_ee$, are deflected sideways by the
magnetic field, and a Hall electric field ${\bf E_H}={\bf v_e}\times
{\bf B}/c$ develops to balance the magnetic force. The variation on
the usual experiment is to allow the conductivity to depend on
height. In this case, $\vec{J}$ and therefore $\vec{v_e}$ and
$\vec{E_H}$ depend on height, giving an emf ${\mathcal E}=\int
\vec{E_H}\vdot\vec{dl}$ which causes a magnetic field component to
grow along the current direction. The result is that the magnetic field lines are sheared by the electron flow. In the star, this shear arises because of
gradients in $\vec{B}$, $\vec{J}$, and $n_e$. The Hall
timescale given in equation (\ref{eq:thall}) is simply the shearing
time of the electrons, $t_{\rm Hall}=L/v_e$.

We now calculate $t_{\rm Hall}$ for densities $\rho$ from $\approx
10^9\ {\rm g\ cm^{-3}}$, through neutron drip at $\rho\approx 4\times 10^{11}\ {\rm g\
cm^{-3}}$, down to the crust/core interface at $\rho\approx 10^{14}\
{\rm g\ cm^{-3}}$. For $\rho\lesssim 6\times 10^{12}\ {\rm g\
cm^{-3}}$, the relativistic, degenerate electrons dominate the
pressure. The electron Fermi energy and pressure are $E_{F,e}=51\ {\rm
MeV}\ \rho_{12}^{1/3}Y_e^{1/3}$, and $P_e=1.2\times 10^{31}\ {\rm erg\
cm^{-3}}\ \rho_{12}^{4/3}Y_e^{4/3}$, where $\rho_{12}=\rho/10^{12}\
{\rm g\ cm^{-3}}$, and $Y_e$ is the number fraction of electrons. The
top of the crust is set by $\Gamma=Z^2e^2/ak_BT=175$ (Potekhin \&
Chabrier 2000), where $a=(3/4\pi n_i)^{1/3}$ is the interion spacing,
giving a density
\begin{equation}
\rho_{\rm top}=8\times 10^7\ {\rm g\ cm^{-3}}\ T_8^3\ \left({Z\over
26}\right)^{-6}\left({A\over 56}\right).
\end{equation}
The top of the crust varies strongly with temperature, from $\rho\sim
10^9\ {\rm g\ cm^{-3}}$ in accreting stars, to $\sim 10^5\ {\rm g\
cm^{-3}}$ in isolated, cooling neutron stars.

For $\rho\gtrsim 6\times 10^{12}\ {\rm g\ cm^{-3}}$, the degenerate,
non-relativistic neutrons dominate the pressure. The neutron Fermi
energy and pressure are $E_{F,n}=15\ {\rm MeV}\
\rho_{14}^{2/3}Y_n^{2/3}(f/0.5)$, and $P_n=6.0\times 10^{32}\ {\rm
erg\ cm^{-3}}\ \rho_{14}^{5/3}Y_n^{5/3}(f/0.5)$. Here, $f$ is a factor
which accounts for the interactions between neutrons. We use the fit
of Mackie \& Baym (1977) for $E_{F,n}$ (following Brown 2000).

\begin{figure}
\epsscale{1.1}\plotone{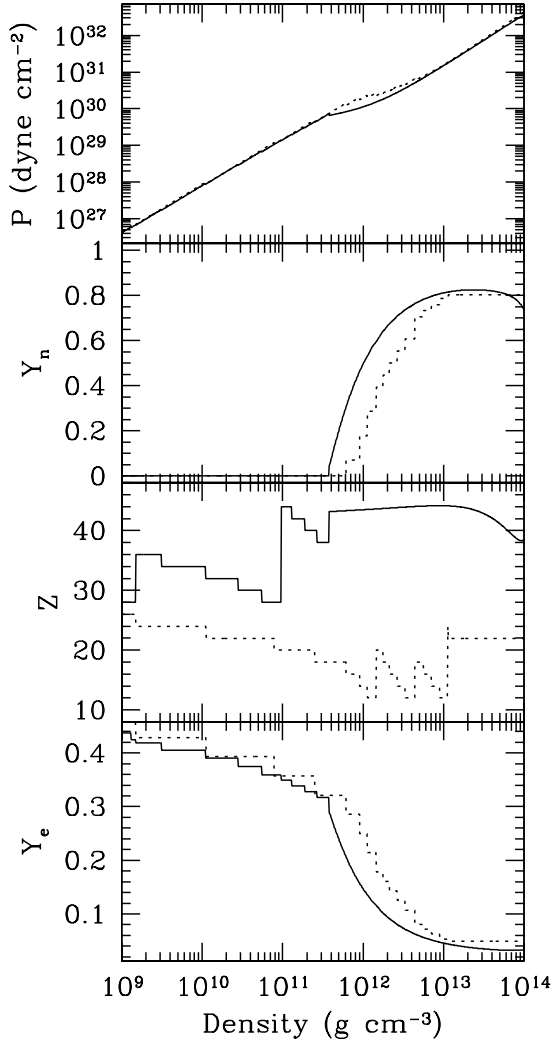}
\caption{Pressure, neutron fraction $Y_n$, nuclear charge $Z$, and
electron fraction $Y_e$ as a function of density for an isolated
crust (solid line) and accreted crust (dotted line).\label{fig:prof}}
\end{figure}

The electron and neutron fractions ($Y_e$ and $Y_n$) are given by the
composition at each depth. An accreting neutron star can replace its
crust with accreted material, in which case the composition is quite
different from an isolated object. Figure \ref{fig:prof} shows the
pressure, $Y_e$, $Y_n$ and nuclear charge $Z$ as a function of depth
in both cases. For an isolated crust, we use the cold-catalysed matter
results of Haensel \& Pichon (1994) ($\rho<4\times 10^{11}\ {\rm g\
cm^{-3}}$), and Douchin \& Haensel (2001) ($\rho>4\times 10^{11}\ {\rm
g\ cm^{-3}}$). For the accreted case, we use the results of Haensel \&
Zdunik (1990, hereafter HZ), who calculated the composition changes
due to electron captures and pycnonuclear reactions as fluid elements
traverse the crust during accretion. HZ stop their calculation at
$\rho=1.25\times 10^{13}\ {\rm g\ cm^{-3}}$; at higher densities, we
fix the composition at their final value, giving $Y_e=0.05$ for
$\rho>1.25\times 10^{13}\ {\rm g\ cm^{-3}}$ in our accreted crust. Our
equation of state agrees well with Negele \& Vautherin (1973).

\begin{figure*}
\epsscale{1.0}\plotone{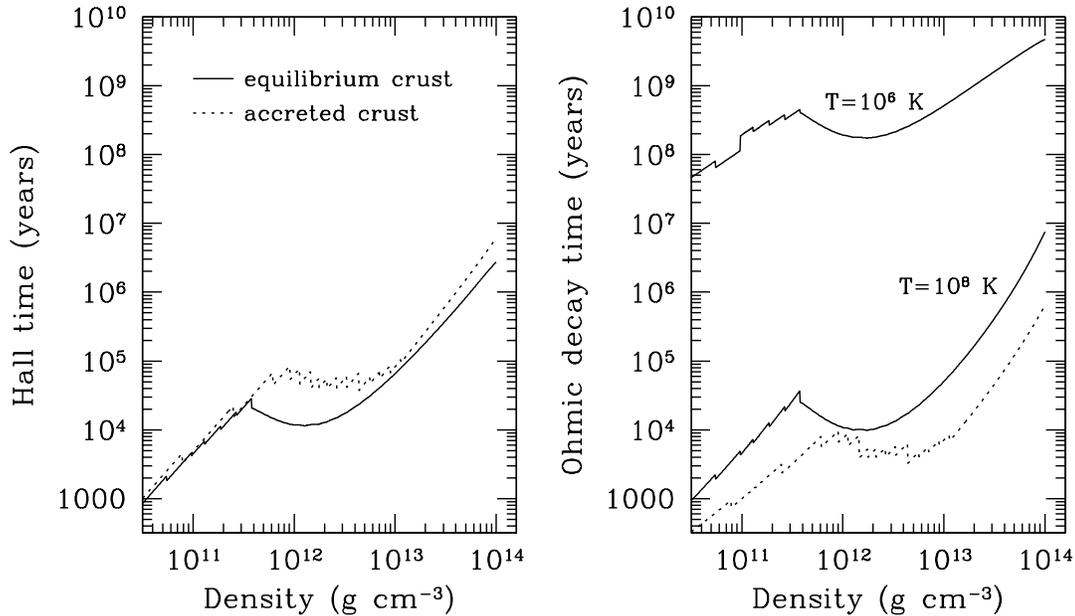}
\caption{Hall timescale $t_{\rm Hall}$ (left panel) and ohmic time $t_{\rm
ohm}$ (right panel) as a function of density for two examples of an
equilibrium crust (solid lines) and accreted crust (dashed line). The
equilibrium crusts have $Q=10^{-3}$ and $T=10^6\ {\rm K}$ and $10^8\
{\rm K}$; the accreted crust has $T=5\times 10^8\ {\rm K}$ and
$Q=1$. We assume the crust is isothermal, with a constant magnetic
field $B=10^{12}\ {\rm G}$ ($t_{\rm Hall} \propto 1/B$), and take the relevant
lengthscale to be the scale height, with local gravity $g=2.45\times
10^{14}\ {\rm cm\ s^{-2}}$.\label{fig:th}}
\end{figure*}

We show the Hall timescale as a function of density for an isolated
and accreted crust in Figure \ref{fig:th}. We choose the lengthscale
$L$ to be the local pressure scale height, $H=P/\rho g$, where $g$ is
the local gravity, assumed constant. For densities between $10^{12}$
and $10^{13}\ {\rm g\ cm^{-3}}$, the Hall time is approximately
constant with increasing density because as neutron drip occurs, the
electron fraction drops dramatically, cancelling the increasing
density and scale height. At lower densities, the electrons set the
pressure, giving $H=77.6\ {\rm m}\
\rho_{12}^{1/3}(Y_e/0.25)^{4/3}(2.45/g_{14})$, and
\begin{equation}
t_{\rm Hall, outer}={5.7\times 10^4\ {\rm yrs}\over B_{12}}\
\rho_{12}^{5/3}\left({Y_e\over 0.25}\right)^{11/3}\left({g_{14}\over
2.45}\right)^{-2}.
\end{equation}
At densities greater than neutron drip, the neutrons dominate the
pressure, giving \newline $H=245\ {\rm m}\ 
\rho_{14}^{2/3}(f/0.5)Y_n^{5/3}(2.45/g_{14})$, and
\begin{equation}\label{eq:thallinner}
t_{\rm Hall, inner}={1.2\times 10^7\ {\rm yrs}\over B_{12}}\
\rho_{14}^{7/3}Y_n^{10/3}\left({Y_e\over 0.05}\right)\left({f\over
0.5}\right)^2 \left({g_{14}\over 2.45}\right)^{-2}.
\end{equation}
We have inserted typical values for $Y_e$ at each density.  In \S \ref{sec:fieldevolution}, we argue that the timescale for overall evolution of the field is roughly the Hall time at the base of the crust, as given by equation
(\ref{eq:thallinner}). The choice $L=H$ introduces the $1/g^2$
dependence of $t_{\rm Hall}$ on the local gravity in the crust, and
therefore some sensitivity to the equation of state of the core.

\subsection{Calculation of the Ohmic decay time}
\label{sec:ohmictime}

\begin{figure}
\epsscale{1.0}\plotone{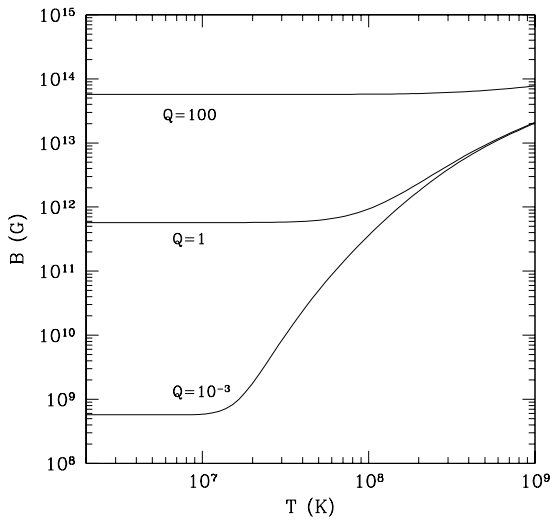}
\caption{Ohmic decay vs.~Hall effect. The solid lines show the $B$
field above which $\Omega\tau \geq 1$ as a function of temperature at a
density of $\rho=10^{14}\ {\rm g\ cm^{-3}}$, and for an isolated
crust.\label{fig:bt}}
\end{figure}
 
For comparison, we now calculate the Ohmic time across a scale height
$t_{\rm Ohm}=4\pi\sigma H^2/c^2$. The conductivity is
$\sigma=n_ee^2/m_\star\nu$, or
\begin{equation} \sigma=1.52\times
10^{25}\ {\rm s^{-1}}\ \left({\nu\over 10^{17}\ {\rm
s^{-1}}}\right)^{-1}\ \rho_{12}^{2/3}Y_e^{2/3},
\end{equation}
where $\nu$ is the electron scattering frequency, and
$m_\star=E_F/c^2\gg m_e$ is the electron effective mass, giving
$t_{\rm ohm}=4\times 10^5\ {\rm yrs}\ \sigma_{25}H_4^2$.

The electrical conductivity is set by electron scattering from phonons
and impurities.  The phonon scattering frequency is
$\nu_{ep}=(13\alpha k_BT\Lambda_{ep}/\hbar)[1+(\Theta/3.5
T)^2]^{-1/2}\exp(-T_U/T)$ (Baiko \& Yakovlev 1995, 1996), where
$\Lambda_{ep}$ is the order unity Coulomb logarithm, and $\Theta_D$ is
the Debye temperature,
\begin{equation}
\Theta_D={0.45\hbar\over k_B}\left({4\pi Z^2e^2n_i\over A_{\rm
cell}m_p}\right)^{1/2}=3.5\times 10^9\ {\rm K}\ Y_e\rho_{12}^{1/2},
\end{equation}
with $A_{\rm cell}$ the number of baryons per unit cell. Except for
very young neutron stars, $T\ll\Theta_D$. The exponential suppression
of the scattering frequency for $T<T_U=8.7\ \times 10^7\ {\rm K}
\rho_{14}^{1/2}(Y_e/0.05)(Z/30)^{1/3}$ occurs because Umklapp
processes freeze out below this temperature. For $T>T_U$, we find
\begin{equation}\label{eq:sigp} 
\sigma_p=\cases{1.2\times 10^{24}\ {\rm s^{-1}}\
(\rho_{12}^{7/6}/T_8^2)(Y_e/0.25)^{5/3}\cr 1.8\times 10^{25}\ {\rm
s^{-1}}\ (\rho_{14}^{7/6}/T_8^2)(Y_e/0.05)^{5/3} \cr}.
\end{equation}
The impurity scattering frequency is (Itoh \& Kohyama 1993),
$\nu_{eQ}=1.75\times 10^{16}\ {\rm s^{-1}}\ xQ\Lambda_{eQ}/Z$, where
$x=p_F/m_ec=100\ \rho_{12}^{1/3}Y_e^{1/3}$, $\Lambda_{eQ}$ is the
order unity Coulomb logarithm, and $Q$ is the impurity factor, giving
\begin{equation}\label{eq:sigQ}
\sigma_Q=\cases{ 1.6\times 10^{25}\
(\rho_{12}^{1/3}/Q)(Y_e/0.25)^{1/3}(Z/30)\cr 4.4\times 10^{25}\
(\rho_{14}^{1/3}/Q)(Y_e/0.05)^{1/3}(Z/30)\cr}.
\end{equation}
For clarity, we set $\Lambda_{ep}=\Lambda_{eQ}=1$ in equations
(\ref{eq:sigp}) and (\ref{eq:sigQ}); in our numerical calculations, we
calculate these factors according to Itoh \& Kohyama (1993) and Baiko
\& Yakovlev (1996), and include the exponential suppression of phonon
scattering for $T<T_U$ as in Gnedin, Yakovlev, \& Potekhin (2001). Our
conductivities agree well with those of Potekhin et al.~(1999).

Impurity scattering dominates either at low temperatures $T<T_U$ (when
phonon scattering is suppressed), or for high impurity levels
$Q\gtrsim Q_{\rm crit}$, where
\begin{equation}\label{eq:Qcrit}
Q_{\rm crit}=\cases{2.4\
(T_8^2/\rho_{14}^{5/6})(Z/30)(0.05/Y_e)^{4/3}\cr 13\
(T_8^2/\rho_{12}^{5/6})(Z/30)(0.25/Y_e)^{4/3}\cr}
\end{equation}
is given by the ratio of equations (\ref{eq:sigp}) and
(\ref{eq:sigQ}). For an equilibrium crust in an isolated neutron star,
Flowers \& Ruderman (1977) estimated $Q\approx 10^{-3}$, in which case
the conductivity is set by phonon scattering above $10^{7-8}\ {\rm
K}$, depending on position in the crust, and by impurity scattering at
lower temperatures. However, recent calculations by Jones (2001) suggest a much larger value $Q\sim 10$, in which case impurity scattering may set the conductivity even for $T>10^8\ \mathrm{K}$.

In accreting neutron stars, the crust temperature
is high ($\gtrsim 10^8\ {\rm K}$) for rapid accretors ($\gtrsim
10^{-11}\ M_\odot\ {\rm yr^{-1}}$), in which case phonons dominate for
$Q\lesssim {\cal O}(1)$. However, $Q$ is set by the composition of the
nuclear burning which occurs at low densities, and is likely to be
$\gtrsim {\cal O}(1)$ (for example, Schatz et al.~1999 found $Q\approx
100$), in which case impurities dominate.

When impurities dominate, the Ohmic time is
\begin{eqnarray}
t_{\rm Ohm}=5.7\ {\rm Myrs}\ {\rho_{14}^{5/3}\over Q}
\left({Z\over 30}\right) 
\left({Y_e\over 0.05}\right)^{1/3}
\left({Y_n\over 0.8}\right)^{10/3}\nonumber\\
\left({f\over 0.5}\right)^{2}
\left({g_{14}\over 2.45}\right)^{-2},
\label{eq:tohmbase_imp}
\end{eqnarray}
and when phonons dominate, it is
\begin{eqnarray}
t_{\rm Ohm}=2.2\ {\rm Myrs}\ {\rho_{14}^{15/6}\over T_8^2}
\left({Y_e\over 0.05}\right)^{5/3}
\left({Y_n\over 0.8}\right)^{10/3}\nonumber\\
\left({f\over 0.5}\right)^{2}
\left({g_{14}\over 2.45}\right)^{-2}.
\label{eq:tohmbase_eph}
\end{eqnarray}
The ohmic time $t_{\rm Ohm}=4\pi\sigma L^2/c^2$ is shown in Figure
\ref{fig:th} for equilibrium crusts with $Q=10^{-3}$, and $T=10^6\ {\rm
K}$ and $10^8\ {\rm K}$, and for an accreted crust with $T=5\times 10^8\ {\rm
K}$ and $Q=1$.

\subsection{When does the Hall Effect dominate Ohmic decay?}

The ratio of Ohmic time $t_{\rm ohm}$ to the Hall time $t_H$
is equivalent to the product of electron cyclotron frequency and
electron collision time, $\Omega\tau=eB\tau/m_\star c$. Large $\Omega\tau$
means that $t_{\rm Hall}\ll t_{\rm Ohm}$, and the Hall effect
dominates. If the conductivity is set by impurities ($Q>Q_{\rm crit}$
or $T<T_U$)
\begin{equation}\label{eq:omegatau1}
\Omega\tau=\cases{ 7\
(B_{12}/Q\rho_{12}^{2/3})(Z/30)(Y_e/0.25)^{-2/3}\cr 1\
(B_{12}/Q\rho_{14}^{2/3})(Z/30)(Y_e/0.05)^{-2/3}\cr },
\end{equation}
and if the conductivity is set by phonons ($Q<Q_{\rm crit}$ and
$T>T_U$)
\begin{equation}
\Omega\tau=\cases{
0.6\ (B_{12}\rho_{12}^{1/6}/T_8^2)(Y_e/0.25)^{2/3} \exp(T_U/T) \cr
0.4\ (B_{12}\rho_{14}^{1/6}/T_8^2)(Y_e/0.05)^{2/3} \exp(T_U/T) \cr
},
\label{eq:omegatau}
\end{equation}
These different regimes are summarized in Figure \ref{fig:bt}.  We
show the $\Omega\tau=1$ contour in the $B$--$T$ plane for conditions
near the base of the crust ($\rho=10^{14}\ {\rm g\ cm^{-3}}$). Note,
however, that $\Omega \tau$ is reasonably insensitive to $\rho$, so
that all depths change from Ohmic decay-dominated to Hall-dominated at
roughly the same time.

Figure \ref{fig:bt} shows very clearly the difference in evolution
between isolated and accreting neutron stars. An isolated neutron star cools to $T<T_U$ in $\approx 1\ \mathrm{Myr}$, after which impurities set the conductivity giving $\Omega\tau \simeq B_{12}/Q$ at the base. The extent to which Hall effects dominate then depends sensitively on the value of $Q$ for an isolated crust. The recent calculation by Jones (2001) of the formation enthalpies of point defects in the crust gives $Q\gtrsim 1$ at densities greater than neutron drip (see also de Blasio 2000 for densities lower than neutron drip). In this case Ohmic decay dominates the evolution even at late times. The original estimates of Flowers \& Ruderman (1977) gave $Q\ll 1$, in which case the Hall effect dominates for $T<T_U$. An accreting neutron star has $T>T_U$ because of heating by accretion (e.g. Brown 2000), and in addition may have a large impurity parameter $Q>1$ because of the complex mixture of nuclei produced by burning of the accreted hydrogen and helium (Schatz et al.~1999). This means that whether impurities or phonons dominate, the conductivity is poor enough that $\Omega\tau\ll 1$, implying that the Hall effect is not important for an accreting star (even after accretion terminates, if the crust is impure).

For an isolated neutron star with a pure crust ($Q\ll 1$), Figure \ref{fig:bt} allows us to estimate when the Hall effect will dominate Ohmic decay. For $Q=10^{-3}$ at
the base of the crust, the critical temperature below which the Hall
effect is dominant is $T\approx 6\times 10^7, 2\times 10^8$, and
$6\times 10^8 {\rm K}$ for magnetic fields $B=10^{11}, 10^{12}$ and
$10^{13}\ {\rm G}$, respectively. At these high temperatures, $T \geq
T_U$ and the conductivity is set by electron-phonon scattering. To
estimate the core temperature as a function of time, we adopt the
cooling law for modified URCA neutrino emission with specific heat
given by normal (unpaired) neutrons, $ t \simeq 10^6\ {\rm yr}\
T_8^{-6}$ (e.g.~Shapiro and Teukolsky 1983). Using 
equation (\ref{eq:omegatau}), the age at
which the Hall effect becomes dominant is
\begin{equation}\label{eq:switch}
t_{\rm switch} \simeq 2\times 10^4\ {\rm yr}\ B_{12}^{-3}.
\end{equation}
Ohmic decay is more important in lower field stars since the Hall
effect takes over at a later time. For typical pulsar magnetic fields,
the Hall effect starts to dominate while the star is still quite
young. Ohmic decay calculations ignoring the Hall effect are only valid for $t<t_\mathrm{switch}$.


\section{ Field Evolution When $\Omega\tau \ll 1$ }
\label{sec:ohmicevolution}

In this section, we compute the Ohmic decay of currents in the crust
at early times when $\Omega\tau\ll 1$, before the Hall effect
dominates the field evolution. We first calculate Ohmic decay modes, and then discuss the Green's function solution, which shows self-similar behavior at late times. We show that these solutions are simply related, giving a general way to understand the distribution of currents in the crust at a given age, as well as the decline of the surface field. This constrains the "initial condition" for the later evolution by the Hall effect.

\begin{figure*}
\epsscale{1.0}\plottwo{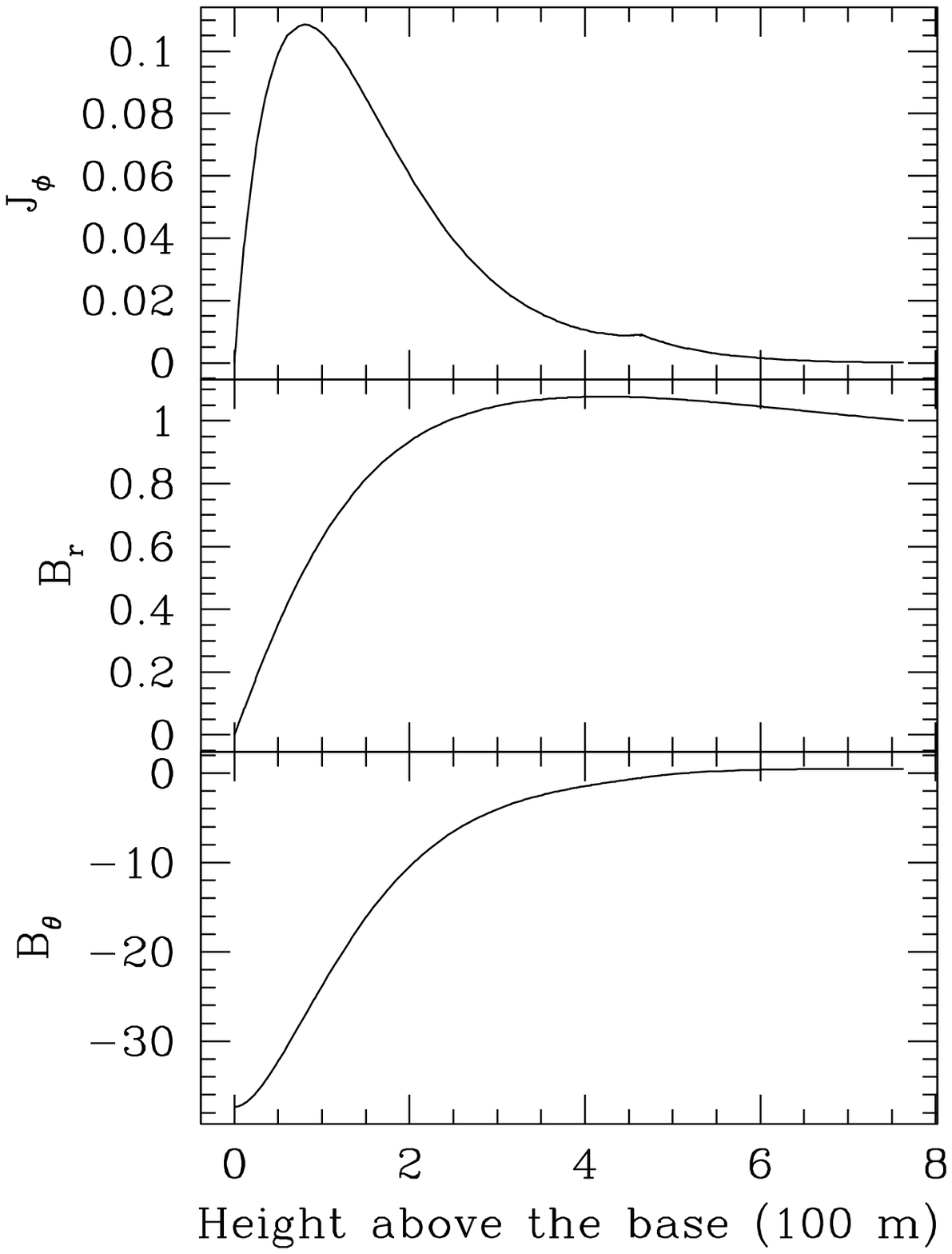}{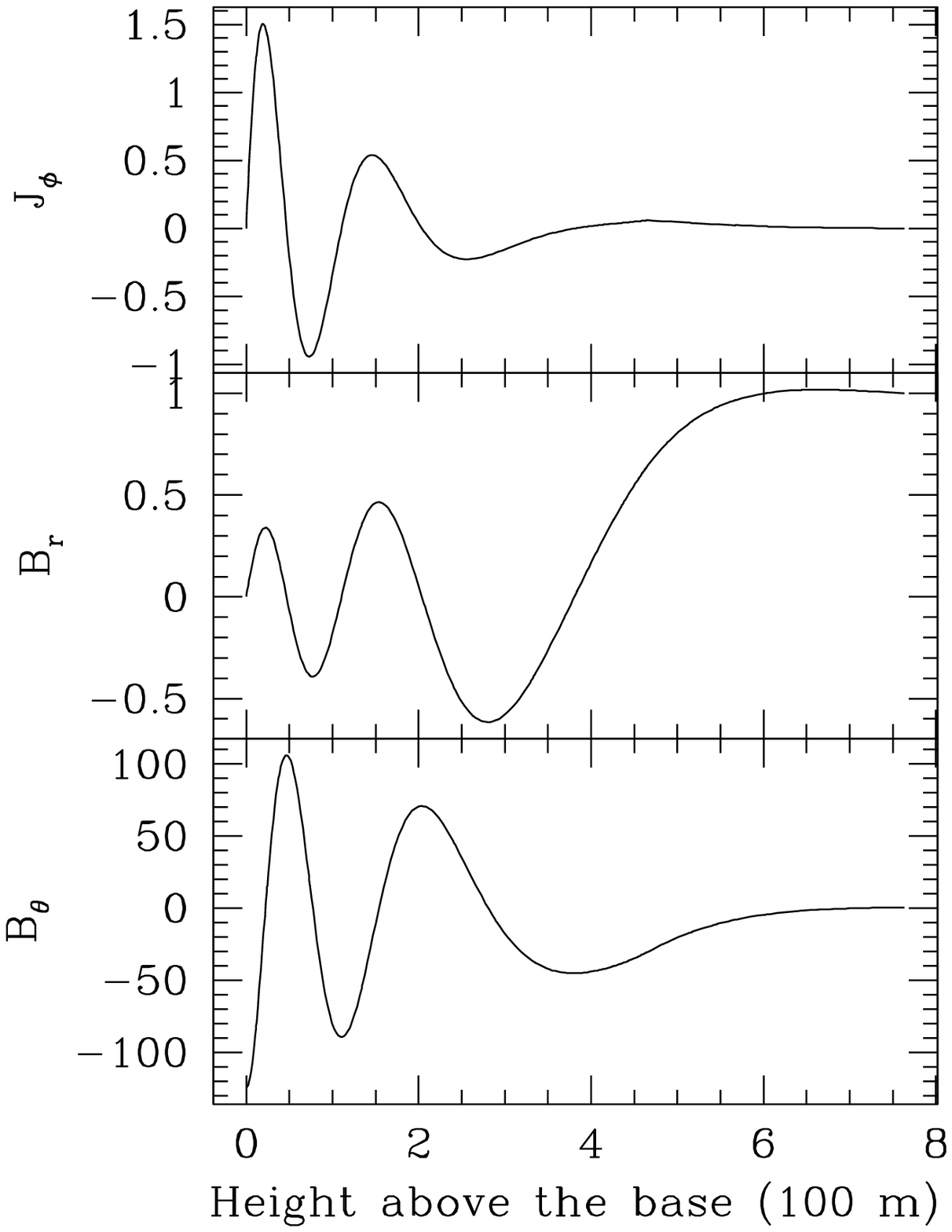}
\caption{Ohmic decay mode eigenfunctions for an equilibrium crust with
$T=10^8\ {\rm K}$ and $Q=10^{-3}$. We show the lowest order mode
($n=1$) in the left panel and the $n=5$ mode in the right
panel.The inner boundary condition is $B_r=0$. \label{fig:mode}}
\end{figure*}

\begin{figure*}
\epsscale{1.0}\plottwo{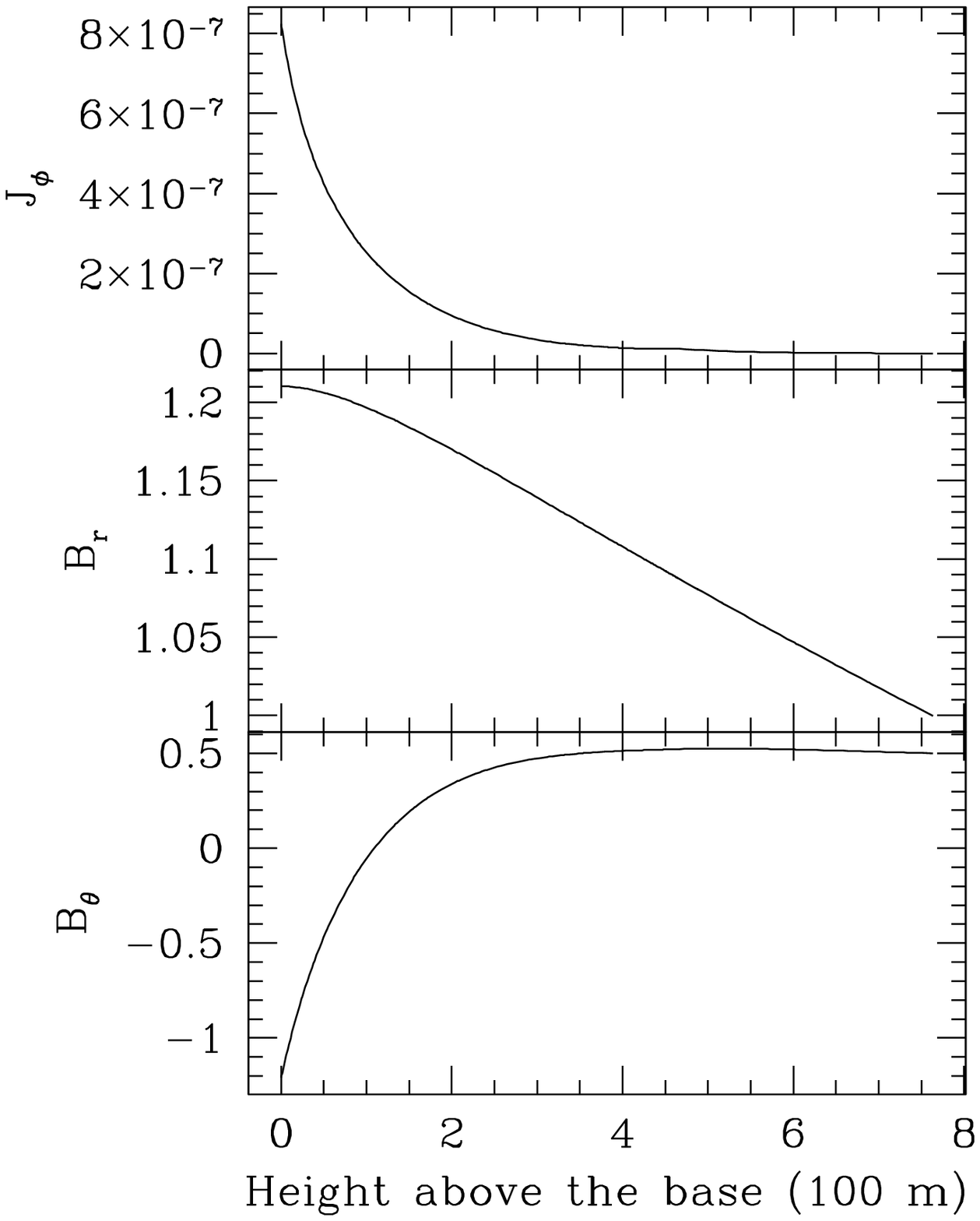}{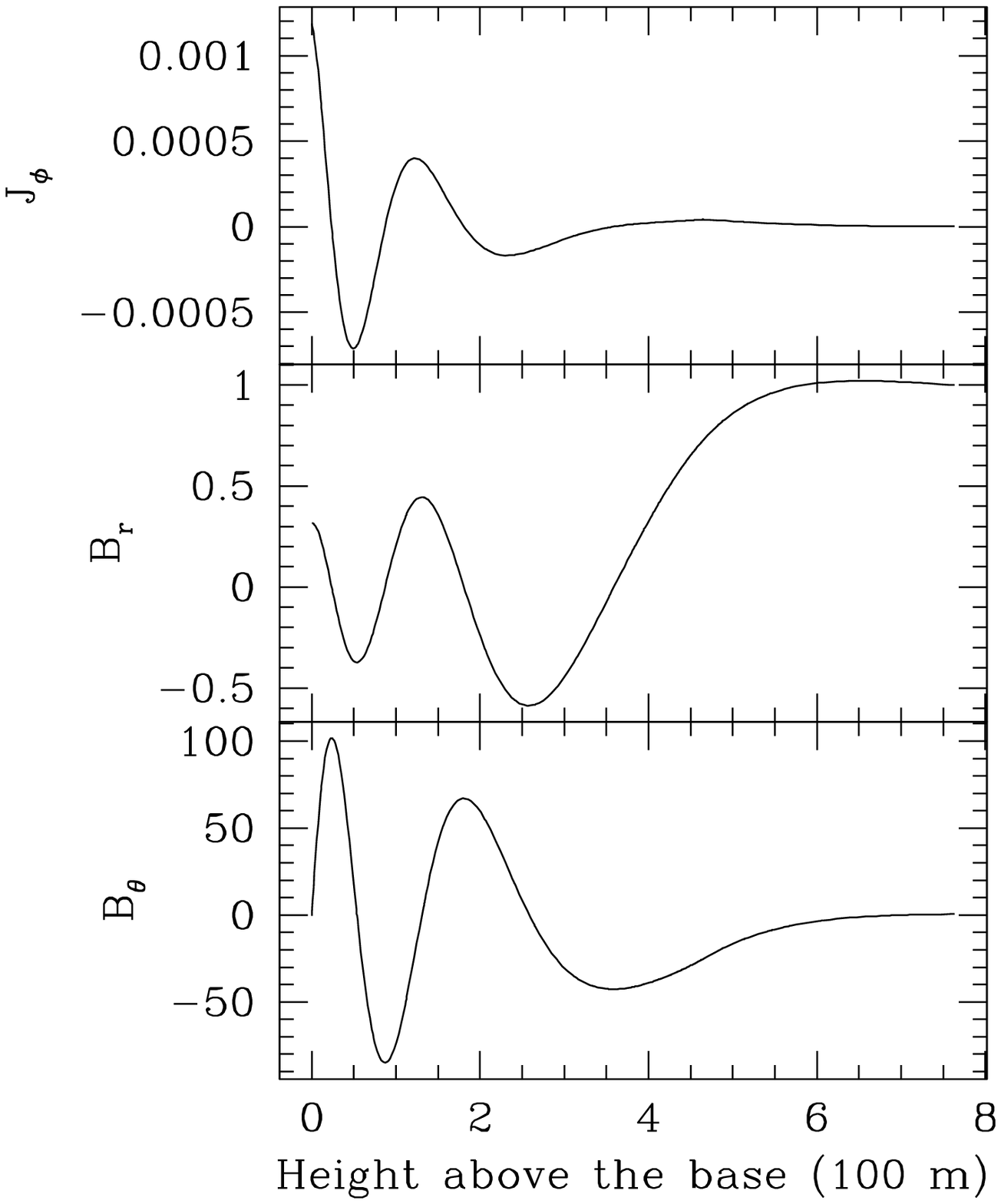}
\caption{Same as Figure \ref{fig:mode}, but with an inner dipole boundary
condition. }
\label{fig:mode2}
\end{figure*}

\subsection{Ohmic decay modes}

We now calculate the Ohmic decay modes for currents confined to the
crust. We write the poloidal magnetic field in terms of the potential
$f(r,t)$, $\vec{B}=\curl \curl [ f(r,t) Y_l^m(\theta,\phi)]$
(Chandrasekhar 1961). The field components are
\begin{equation}
\vec{B}={l(l+1)\over r^2}fY_l^m \vec{e}_r + {1\over r}{df\over dr}
\left( {\partial Y_l^m \over\partial\theta} \vec{e}_\theta + {\partial
Y_l^m \over\partial\phi} \vec{e}_\phi\right).
\end{equation}
Faraday's law (eq.~[\ref{eq:faraday}]) with $\vec{E}=\vec{J}/\sigma$
gives
\begin{equation}
{\partial f\over\partial t} = \eta(r,t)\left({\partial^2f\over
\partial r^2}-{l(l+1)\over r^2}f\right),
\label{eq:poloidaleqn}
\end{equation}
where $\eta=c^2/4\pi\sigma$ is the magnetic diffusivity.

When Ohmic decay determines the evolution of the field, the
conductivity is mainly set by phonons, in which case the time-dependence of
$\eta$ arises only through the temperature, $\eta\propto
T^2$. Following Urpin et al.~(1994), we absorb this time-dependence
into a new time variable $\hat t$, given by $d\hat t\propto T^2\
dt$. For modified URCA cooling $T\propto t^{-1/6}$,
we find $t=(\hat t/1.5\ {\rm Myr})^{3/2}\ {\rm Myr}$ when $\eta$ is
evaluated at the reference temperature $T_8=1$. We then look for Ohmic
decay modes, $f\propto \exp(-\hat t/\tau)$, giving
\begin{equation}\label{eq:modes}
{d^2f\over dr^2}-{l(l+1)\over r^2}f=-{f\over\eta\tau}.
\end{equation}
Note that exponential decay in $\hat t$ represents a slower decay in
real time $t$, because the conductivity of the cooling crust grows
with time, reducing the Ohmic dissipation rate.

Equation (\ref{eq:modes}) is an eigenvalue problem for $\tau$ given
boundary conditions for $f$. At the top of the crust, we match onto a
vacuum field, giving $f\propto r^{-l}$, and normalize the field such
that $B_r=1$ there. At the base of the crust, we consider two
different boundary conditions, either $B_r=0$ at the base, implying
the field is confined entirely to the crust, or we assume the field
connects to an interior vacuum field, an example of a field which
threads the neutron star core\footnote{In fact, if the field
penetrates the core, then there is no strict boundary condition
quantizing the phase in units of roughly $n\pi$ in the crust. The
vertical wavelength in the core will be much smaller than in the
crust, and then the value and slope at the crust core interface will
reflect quantizing the phase in the core. For eigenmodes with many
nodes in the crust, this will not change the waves significantly, but
for long decay times $\gg 10^6\ {\rm yr}$, this boundary condition is
crucial.}. 

A WKB analysis clarifies the basic properties of the modes. Writing
$f\propto\exp(i\int^r dr k_r)\exp(-\hat{t}/\tau)$, we find the
dispersion relation
\begin{equation} 
k_r^2={1\over \eta\tau}-{l(l+1)\over r^2} \simeq
\left(\eta\tau\right)^{-1}
\label{eq:ohmicwkb}
\end{equation}
where $k_r$ is the radial wavenumber and $\tau$ is the decay time. We
ignore the horizontal wavenumber term, which is negligible unless $l
\geq R/H_b \simeq 20$. The WKB solution, including amplitude and
phase, is
\begin{equation}\label{eq:wkbsolution}
f\propto \eta^{1/4} \exp\left( \pm i \int^r
{dr\over\left(\eta\tau\right)^{1/2}}\right).
\end{equation}
For $f$, the amplitude and wavelength grow $\propto\eta^{1/2}$ toward
the top of the crust. The magnetic field and current components have
the same oscillatory phase, but their envelopes scale as $B_r \propto
\eta^{1/4}$, $B_{\theta,\phi} \propto \eta^{-1/4}$, and
$J_{\theta,\phi} \propto \eta^{-3/4}$. A WKB turning point is reached
at the depth $z_{tp}(\tau)$ at which $k_z H_\sigma \sim 1$, or
equivalently, where the mode decay time equals the local diffusion
time. Above this depth, the left and right sides of equation
(\ref{eq:modes}) can no longer balance, leading to evanescent
behavior.

We show two examples of Ohmic decay eigenmodes with $l=1$ in Figures
\ref{fig:mode} ($B_r=0$ at the base) and \ref{fig:mode2} (vacuum field
at the base). The left panel in each case shows the lowest order
eigenmode. When $B_r=0$ at the base, $B_r$ varies rapidly with depth,
giving $B_\theta\approx (R/H)B_r\gg B_r$ at the base; for a vacuum boundary condition at the base, $B_r$ varies only slightly across the crust, giving $B_r \sim
B_\theta$ at the base. For short wavelength modes (right panels), the
eigenfunctions become insensitive to the lower boundary condition; the
effect of the boundary condition is to introduce an overall phase
shift. The WKB amplitude scalings and turning point can be seen
clearly for these modes.

Table \ref{tab:modes} gives Ohmic decay times for numbers of nodes
$n=1-5$, different temperatures, impurity fractions and inner boundary
conditions. The results are scaled in terms of the Ohmic time at the
base, which is listed in the table.  For $T=10^8\ K$, the decay times
for the lowest order modes are $\tau_1=2.1\ \times 10^6\ {\rm yr}$ and
$\tau_1=1.0 \ \times 10^8\ {\rm yr}$ for the $B_r=0$ and vacuum inner
boundary conditions. As pointed out by Pethick \& Sahrling (1995), the
difference between these two cases is that the lowest order mode has a
decay time $\approx R/H$ times larger when the field penetrates into
the core, because the lengthscale on which $B_r$ changes through the
crust is the radius instead of the crust thickness.

\begin{deluxetable}{ccc}
\tablewidth{0pt}
\tablecaption{Ohmic Decay Modes (dipole)\label{tab:modes}}
\tablehead{
\colhead{$n$} & \colhead{$t_{\rm Ohm}/\tau_n$} & \colhead{$t_{\rm Ohm}/n^2\tau_n$}
}
\startdata
\multicolumn{3}{c}{$T_8=1$, $Q=10^{-3}$, $t_{\rm Ohm}=7.5\times 10^6\ {\rm yrs}$}\\
\hline
1 & 3.60 & 3.60\\
2 & 18.2 & 4.56\\
3 & 44.0 & 4.89\\
4 & 82.0 & 5.12\\
5 & 133.0 & 5.33\\
\hline
\multicolumn{3}{c}{$T_8=3$, $Q=10^{-3}$, $t_{\rm Ohm}=7.0\times 10^5\ {\rm yrs}$}\\
\hline
1 & 2.32 & 2.32\\
2 & 12.1 & 3.03\\
3 & 29.8 & 3.31\\
4 & 55.8 & 3.49\\
5 & 90.4 & 3.61\\
\hline
\multicolumn{3}{c}{$T_8=0.01$, $Q=10^{-3}$, $t_{\rm Ohm}=4.7\times 10^9\ {\rm yrs}$}\\
\hline
1 & 0.315 & 0.315\\
2 & 2.17 & 0.542\\
3 & 5.67 & 0.630\\
4 & 10.8 & 0.675\\
5 & 17.6 & 0.703\\
\hline
\multicolumn{3}{c}{$T_8=1$, $Q=10^{-3}$, $t_{\rm Ohm}=7.5\times 10^6\ {\rm yrs}$}\\\multicolumn{3}{c}{inner dipole boundary condition}\\
\hline
1 & 0.0729 & 0.0729\\
2 & 8.76 & 2.19\\
3 & 28.8 & 3.20\\
4 & 60.5 & 3.78\\
5 & 105 & 4.20\\
\enddata
\end{deluxetable}

\begin{figure}
\epsscale{1.0}\plotone{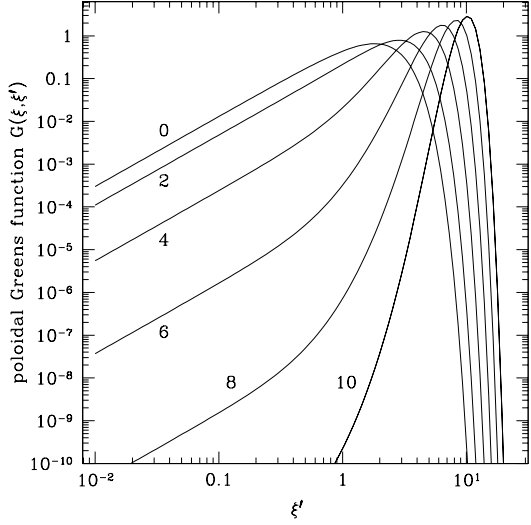}
\caption{ Green's function for Ohmic decay of poloidal magnetic field.
The conductivity profile is $\alpha=7/2$. Each line is labelled with
the value of $\xi$. }
\label{fig:polGf}
\end{figure}

\subsection{Green's function solution}
\label{sec:green}

An alternative method of solving the diffusion equation is to find the Green's function. We now calculate the eigenmodes and Green's function for a plane parallel crust with power-law diffusivity $\eta\propto z^{-\alpha}$ (where $z$ is the depth).\footnote{Although we choose a power law diffusivity mainly because it allows an analytic solution, this is also a good approximation since the conductivity is a power law in both position and time over a large range of scales.}. We start by defining the similarity variable
\begin{equation}
\xi(z,\hat t)=\int_0^z {dz\over \sqrt{\eta \hat t}}.
\label{eq:xi}
\end{equation}
Useful quantities are $\hat t_\mathrm{diff}(z)=\xi^2\hat t\propto z^{\alpha+2}$, the characteristic diffusion time to depth $z$, and $z_\mathrm{diff}(\hat t)\propto \hat t^{1/(\alpha+2)}$, the depth for which the diffusion time is the current age, $\hat t=\hat t_\mathrm{diff}$, given by $\xi(z_\mathrm{diff},\hat t)=1$.

Setting $\eta\propto z^{-\alpha}$, we look for solutions $\propto f(\xi)\exp(-\hat t/\tau)$, where $\xi=\xi(z,\tau)$, giving
\begin{equation}
f(\xi)=\xi^\nu J_{-\nu}(\xi)
\label{eq:poloidalbessel}
\end{equation}
where $\nu=(\alpha+2)^{-1}$, and $J_{-\nu}$ is a Bessel function of the first kind. The solution with order $-\nu$ becomes constant near the surface, therefore matching onto a vacuum field at low densities. The decay time $\tau$ for each mode is determined by enforcing the boundary condition at the base. These eigenmodes have similar properties to those determined in \S 3.1. The turning point for each mode occurs at $\xi \sim 1$, or depth $z_\mathrm{diff}(\tau)$. Since the amplitude of $J_\nu(\xi)$ is $\propto \xi^{-1/2}$ for $\xi\gg 1$, we find $f\propto z^{-\alpha/4}\propto \eta^{1/4}$ at large depths, as expected from the WKB scalings (eq.~[\ref{eq:wkbsolution}]).

The Green's function is obtained by writing an initial delta function
as a sum over the eigenmodes given by equation (\ref{eq:poloidalbessel}), and using the Fourier-Bessel inversion
formula (Jackson 1975, pg.~100) and a standard integral over Bessel
functions (Watson 1966, pg.~395). We find the Green's function
solution (see also Eichler \& Cheng 1989)
\begin{equation}\label{eq:greensolution}
f(\xi,\hat{t})=\int d\ln \xi\prime\ G(\xi,\xi\prime) f_0(z\prime)
\end{equation}
where
\begin{equation}\label{eq:green}
G(\xi,\xi\prime) = \frac{1}{2}\xi^\nu (\xi\prime)^{2-\nu} I_{-\nu}\left(\frac{\xi \xi\prime}{2} \right) \exp \left(-
\frac{\xi^2+\xi\prime^2}{4} \right),
\label{eq:greensfunction}
\end{equation}
$\xi=\xi(z,\hat{t})$, $\xi\prime=\xi(z\prime,\hat{t})$, $I_\nu$ is
a modified Bessel function, and $f_0(z)$ is the initial field profile. For $\xi\ll 1$, $G$ becomes independent of $\xi$,
\begin{equation}\label{eq:approxG}
G\propto \left(\xi^\prime\right)^{2(1-\nu)}\exp\left(-{\xi^{\prime 2}\over 4}\right)
\end{equation}
Figure \ref{fig:polGf} shows the Green's function for $\alpha=7/2$ and $\nu=2/11$, appropriate for electron-phonon conductivity above neutron drip. 

The evolution in time can be understood by taking various limits of equation (\ref{eq:greensolution}). Consider the case where the currents extend down to a finite depth $z_{\rm init}$ initially, $f_0(z>z_\mathrm{init})=0$. For early times (or sufficiently large depths), the field is unchanged: $G\approx \delta(\xi^\prime-\xi)$ for $\xi\gg 1$, giving
\begin{equation}
f(z,\hat{t})\simeq f_0(z)\hspace{1 cm}\xi\gg 1,\ z>z_\mathrm{diff}(\hat t)
\end{equation}
For late times ($\xi\ll 1$), equation (\ref{eq:approxG}) applies, and the solution is {\em independent of depth}, with a time-dependence determined by the integral
over $\xi^\prime$.  When $z_\mathrm{diff}(\hat t)<z_\mathrm{init}$, the peak of the Green's function at $\xi^\prime\approx 1$ gives
\begin{equation}\label{eq:constantphase}
f(z,\hat{t}) \simeq f_0(z_{\rm diff}(\hat{t}))
\hspace{1 cm}\xi\ll 1,\ z_{\rm diff}(\hat t)<z_{\rm init},
\end{equation}
so that the solution is determined by the initial value of the field at the current diffusion depth $z_{\rm diff} (\hat{t})$. If $f_0$ varies slowly with depth, then $f$ will vary slowly with time during this phase, particularly since $z_{\rm diff}$ is a weak power of $\hat{t}$. 
When $z_{\rm diff}(\hat t)>z_\mathrm{init}$, however, the integrand drops to zero at a value of $\xi^\prime<1$, so that equation (\ref{eq:approxG}) gives $G\propto (\xi^\prime)^{2(1-\nu)}$. Writing the solution as an integral over $z^\prime$, the time-dependence factors out as
\begin{eqnarray}\label{eq:bdecay}
f(z,\hat{t})\propto \hat{t}^{\nu-1} \propto\hat{t}^{-(\alpha+1)/(\alpha+2)}
\hspace{0.5 cm}\xi\ll 1,\ z_\mathrm{diff}(\hat t)>z_{\rm init}.
\end{eqnarray}
Therefore at late times, the field decays as a power law in time, independent of the 
initial field profile (Sang \& Chanmugam 1987). Since $\hat t\propto t^{2/3}$, we find that the surface field decays as 
\begin{equation}\label{eq:fielddecay}
\frac{B(t)}{B_0}=\left( \frac{t_{\rm diff}(z_{\rm init})}{t}\right)^{\delta},
\end{equation}
with $\delta=0.55$ ($\alpha=7/2$) for $z_\mathrm{init}$ above neutron drip, and $\delta=0.49$ ($\alpha=7/4$) for $z_\mathrm{init}$ below neutron drip. For completeness, we also note that when impurity scattering dominates the conductivity, $\hat t\propto t$ and $\alpha\simeq 4/3$, giving $\delta=0.7$.

The connection between the eigenmode and Green's function solutions is now
easily seen. The Green's function solution implies that $f$ is constant
for $z<z_{\rm diff}(\hat{t})$, and unchanged for $z>z_{\rm diff}(\hat
t)$. In terms of Ohmic modes, at a given time $\hat t$, all modes
with decay times $\tau\ll\hat{t}$, which propagate closer to the
surface than $z_{\rm diff}(\hat{t})$, have exponentially decayed
away. Therefore for $z<z_{\rm diff}(\hat{t})$ the solution is
dominated by the mode with $\tau=\hat{t}$ which has turning point at
$z_{\rm diff}(\tau)$. The power law decay of the surface field at late times arises
because the decay time of the mode which dominates the surface field
increases with time (Sang \& Chanmugam 1987). For $z>z_{\rm diff}(\hat{t})$, the field profile is a superposition of all WKB modes with decay times $\tau \geq \hat{t}$.

\begin{figure}
\epsscale{1.0}\plotone{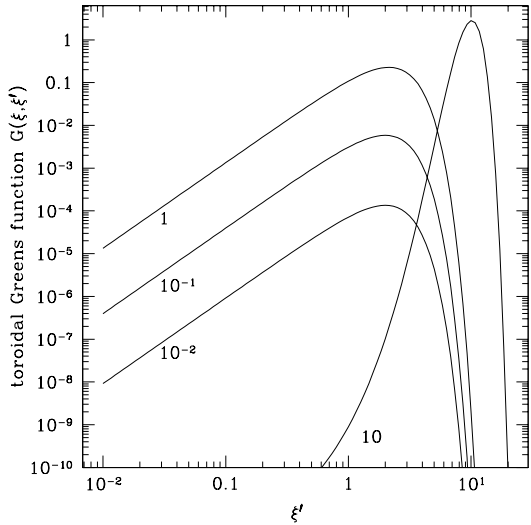}
\caption{ Green's function for Ohmic decay of toroidal magnetic field.
The conductivity profile is $\alpha=7/2$.Each line is labelled with
the value of $\xi$. }
\label{fig:torGf}
\end{figure}

\subsection{Toroidal fields}

Toroidal fields behave in a qualitatively similar way to poloidal
fields, but with some important quantitative differences. Defining the
toroidal field as $\vec{B}=\curl(g(r)Y_l^m\vec{e}_r)$ (Chandrasekhar
1961), we find the diffusion equation is
\begin{equation} 
\frac{\partial g}{\partial t}=\frac{\partial}{\partial r} \left( \eta
{\partial g\over \partial r} \right) - \eta {l(l+1)\over r^2}g,
\end{equation}
with WKB solution
\begin{equation} 
g\propto\eta^{-1/4} \exp\left( \pm i \int^z
{dz\over (\eta \tau)^{1/2}} \right),
\end{equation}
giving $B_{\theta,\phi} \propto \eta^{-1/4}$, $J_{\theta,\phi} \propto
\eta^{-3/4}$, and $J_r \propto \eta^{-1/4}$.  These scalings are
similar to the poloidal case.  For diffusivity $\eta \propto
z^{-\alpha}$, the power-law solution has the same form as in equation
(\ref{eq:poloidalbessel}), but with $\nu=(\alpha+1)/(\alpha+2)$, and the Bessel function must have order $+\nu$ rather than $-\nu$ to ensure zero field at the boundary. The
toroidal Green's function is plotted in Figure \ref{fig:torGf}. The major difference from the poloidal case is for late times, for which the field decays $\propto
\hat t^{-(2\alpha+3)/(\alpha+2)}\propto t^{-1.2}$ (for $\alpha=7/2$), much faster than the poloidal case (compare
eq.~[\ref{eq:bdecay}]). In addition, the field decreases rapidly
towards the surface, $g\propto\xi^{2\nu}\propto z^{\alpha+1}$, implying the interior
toroidal fields are much larger than those near the surface.

\begin{figure}
\epsscale{1.0}\plotone{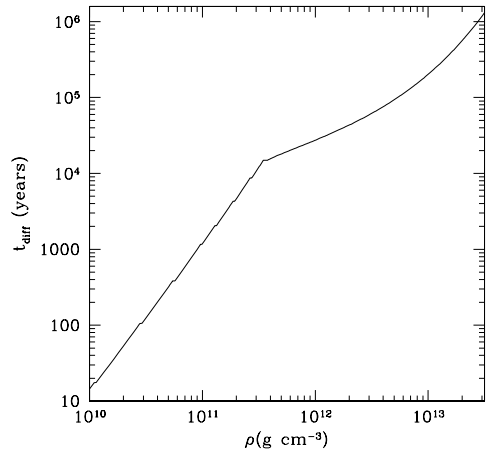}
\caption{ Ohmic time corresponding to the upper turning point as
a function of distance from the top of the crust. We assume phonon conductivity and modified URCA cooling ($T\propto t^{-1/6}$) (for $t\gtrsim 1$ Myr, impurity scattering dominates the conductivity, in which case the Ohmic evolution slows considerably for $Q\ll 1$). (i) For a given age $t$, this plot shows the density at which currents are just beginning to evolve due to Ohmic diffusion. At lower densities, the field is roughly constant with depth, dominated by the Ohmic decay mode with $\tau\approx t$; at higher densities, the field has not yet evolved significantly, and is dominated by Ohmic decay modes with $\tau>t$. (ii) For currents initially distributed up to a density $\rho_\mathrm{init}$, this plot gives the time at which the surface field evolution changes from roughly constant with time to a power law $\propto t^{-0.55}$. }
\label{fig:ohmictime_vs_depth}
\end{figure}

\subsection{Summary of Ohmic decay}

The results of this section give a very simple way to understand the
distribution of currents and the decay of the surface dipole field given the age of the star. For a general conductivity profile $\eta(z)$, the rescaled time corresponding to the
upper turning point is
\begin{equation}
\hat{t}_{\rm diff}(z)=\left( \int_0^z {dz\over\eta^{1/2}(z)}\right)^2,
\end{equation}
where $z=0$ is at the top of the crust. We plot the physical time
$t_\mathrm{diff}\propto \hat t_\mathrm{diff}^{3/2}$ for a detailed crust model with electron-phonon scattering in Figure \ref{fig:ohmictime_vs_depth}.  This plot can be
read as follows: since we have included the fact that the temperature
is decreasing in time, set the ordinate $t_{\rm diff}$ equal to the
age $t$, and then read off the density at which $z=z_{\rm diff}$. At
lower densities, the currents have decayed according to equation (\ref{eq:fielddecay}),
 and we expect $f$
approximately constant, dominated by the Ohmic decay mode with
$\tau\approx t$; at higher densities, the initial field has not yet
evolved significantly, and is dominated by modes with $\tau>t$.
We emphasise again that we assume phonons dominate the conductivity in Figure \ref{fig:ohmictime_vs_depth}. This is a good assumption even if the crust is impure $Q\gtrsim 1$, except at the very base of the crust (see eq.~[\ref{eq:Qcrit}]). After $\approx 1$ Myr, the crust conductivity becomes impurity scattering dominated, and the field decay slows dramatically if $Q\ll 1$.

Our results compare well with the time-dependent calculations of Urpin and collaborators (Urpin \& Muslimov 1992; Urpin \& Konenkov 1997; Page et al.~2000). Their initial profile $f_0(z)$ varies slowly with depth in the crust, giving very little decay of the field for $t<t_{\rm diff}(z_{\rm init})$ (eq.~[\ref{eq:constantphase}]). For $t>t_{\rm diff}(z_{\rm init})$, the decay of the observable dipole field agrees well with equation (\ref{eq:fielddecay}). We do not reproduce the much longer ($\sim 10^8$ year) decay timescales found by Chamugam \& Sang (1987); the reason for this discrepancy is not clear. 

At $t=t_{\rm switch}$ (eq.~[\ref{eq:switch}]) when the Hall effect begins to dominate Ohmic decay, we find that the currents must reside at a
density $\gtrsim 10^{10}\ {\rm g\ cm^{-3}}$ (towards the
top of the crust) for $B=10^{13}\ {\rm G}$, $\gtrsim 10^{12}\ {\rm g\ cm^{-3}}$ (near neutron drip) for $B=10^{12}\ {\rm G}$, and $\gtrsim 10^{14}\ {\rm
g\ cm^{-3}}$ (near the base of the crust) for $B=10^{11}\ {\rm G}$.
Therefore the distribution of currents when the Hall effect begins to dominate
the evolution depends sensitively on the field strength.


\section{Field Evolution When $\Omega\tau \gg 1$ }
\label{sec:fieldevolution}

In this section, we first give a qualitative discussion of how the Hall effect acts on the
magnetic field. We discuss how an initial large lengthscale
poloidal dipole field generates higher wavenumbers, and discuss an interesting time-independent solution of the Hall equation. Next, we present short lengthscale Hall wave solutions for realistic crust models, including a first calculation of the elastic response of the crust to the wave. We discuss the Hall drift instability found by Rheinhardt \& Geppert (2002), and finally discuss the implications of our results for GR's non-linear Hall cascade.

\subsection{ Qualitative discussion of Hall evolution}
\label{sec:qualitative}

In terms of the electron velocity $\vec{v}_e=-\vec{J}/n_e e$, the
induction equation (\ref{eq:inducthall}) is
\begin{equation}\label{eq:qual1}
{\partial\vec{B}\over\partial t}=\curl\left(\vec{v}_e\vcross\vec{B}\right).
\end{equation}
Although this shows a {\it formal} relation to the usual MHD advection
equation, it differs because the electron velocity depends directly on
$\vec{B}$. Nonetheless, it is useful to think of the field moving with the electron fluid
in the same way as the field is tied to fluid elements in MHD. It is instructive to rewrite equation (\ref{eq:qual1}) as
\begin{equation}\label{eq:halladvection}
{D\vec{B}\over Dt} = \vec{B}\cdot\grad\vec{v}_e -
\vec{B}{v_{e,r}\over H_e}
\end{equation}
where $D/Dt=\partial/\partial t+\vec{v_e}\cdot\grad$ is an advective
derivative following the electrons, $v_{e,r}$ is the radial component
of the electron velocity, and $H_e$ is the electron density scale
height. The first term on the right hand side of equation (\ref{eq:halladvection}) represents shearing of
the field by the electron flow; the second term arises because
$\divr\vec{v_e}$ is non-zero for a vertical current because of the
background electron density gradient: $\divr\vec{J}=0$ gives $\divr\vec{v_e}=-\vec{v_e}\cdot\grad n_e/n_e$. For example, as currents flow to deeper regions, the electron density increases, the electron velocity decreases, and the field is amplified by compression.

\begin{figure}
\epsscale{1.1}\plotone{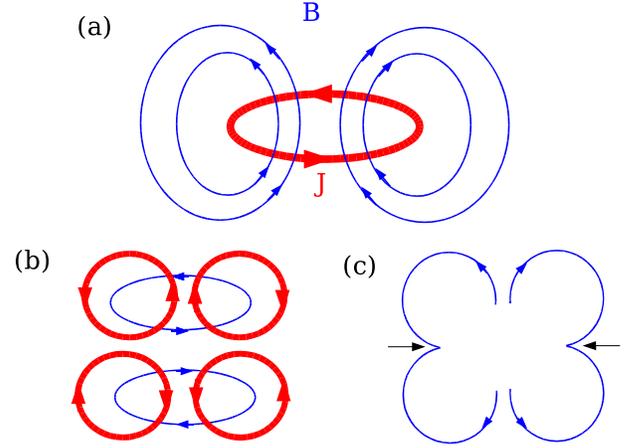}
\caption{The first steps in the evolution of an initial dipole field
by the Hall effect. (a) The toroidal current supporting the dipole
field shears the field at mid-latitudes. (b) The resulting toroidal
field and associated poloidal currents. The net effect is to ``twist''
the original dipole field. (c) The poloidal currents ``pinch'' the
original poloidal dipole field, generating a poloidal quadrupole.}
\label{fig:twist}
\end{figure}

As a first example, consider an axisymmetric poloidal field
$\vec{B_P}=B_r(r,\theta)\vec{e}_r +
\vec{B}_\theta(r,\theta)\vec{e}_\theta$, supported by toroidal
currents which we write in terms of the electron angular velocity
$\Omega_e$ as $\vec{v}_e=\Omega_e(r,\theta) r \sin\theta
\vec{e}_\phi$. In this case, equation (\ref{eq:qual1}) gives
$\partial\vec{B_P}/\partial t=0$, and
\begin{equation}\label{eq:poloidalevol}
{\partial B_\phi\over \partial t} = r\sin\theta
\left(\vec{B_P}\cdot\grad\right)\Omega_e,
\end{equation}
showing that any differential rotation of the electron flow along the
poloidal field lines shears the field, generating a toroidal component
and a net twist. This is a direct analog of the winding up of field
lines by differential rotation (e.g.~Mestel 1999). Figure \ref{fig:twist}(a) shows an initial dipole field and associated
toroidal current. In this case, shear in the electron velocity at
mid-latitudes generates a toroidal field shown in Figure
\ref{fig:twist}(b). The net effect is to twist the original field in
one direction in the northern hemisphere and the opposite direction in
the southern hemisphere. The subsequent evolution is determined by the
{\it poloidal} currents associated with $B_\phi$. They act back on the
original poloidal field, ``pinching'' it at the equator, and
generating a quadrupole component of the poloidal field, as shown in
Figure \ref{fig:twist}(c). In this way, the process continues, and the
smaller and smaller scale components of the field develop.

As a second example, we consider an axisymmetric toroidal field
$\vec{B_T}=B_\phi(r,\theta)\vec{e}_\phi$, supported by poloidal
currents $\vec{v}_e = v_r(r,\theta) \vec{e}_r + v_\theta(r,\theta)
\vec{e}_\theta$. In this case, equation (\ref{eq:qual1}) gives
\begin{equation}
{\partial B_\phi\over\partial t}=-\vec{v}_e\cdot\grad B_\phi
-B_\phi{v_{e,r}\over H_e}+B_\phi\left({v_{e,r}\over r}+ 
{v_{e,\theta}\cot\theta\over r}\right),
\label{eq:toroidalevol}
\end{equation}
and $\partial\vec{B_P}/\partial t=0$, so that an initially toroidal field remains toroidal. The terms on the right hand side of equation (\ref{eq:toroidalevol}) describe advection by poloidal currents, divergence of the velocity flow due to the electron density gradient, and spherical geometry effects. Vainshtein et al.~(2000) solved equation (\ref{eq:toroidalevol}) for a plane-parallel geometry, in which case it reduces to Burger's equation, since only the term involving $H_e$ survives. For a toroidal field with opposite signs in the northern and southern hemispheres, as in Figure \ref{fig:twist}(b), they showed that the toroidal field strength grows at the equator, where the electrons flow inwards to higher density, and is reduced in strength towards the pole, where the electrons flow outwards to lower density (or vice versa depending on the overall field direction). They argued that if the field propagates towards the equator, a thin current sheet will form, rapidly dissipating the magnetic energy. It is not clear whether this effect persists in spherical geometry and including a poloidal component of the field.

These examples show that the field geometry can give a large
correction to the Hall timescale estimate of \S
\ref{sec:halltime}. For example, in Figure \ref{fig:twist}, the
initial evolution of the field is by shearing $B_r$ at
mid-latitudes. If the magnetic field is confined to the crust, we
expect $B_r\approx B_\theta (H/R)$ to vary across a lengthscale $H$,
and $J_\phi\approx (c/4\pi) (B_\theta/H)$. This gives $t_{\rm
Hall}=4\pi n_eeH^2/cB_\theta$, so that the horizontal field is the
appropriate field to use in the Hall time estimates of \S
\ref{sec:halltime}. If the magnetic field penetrates into the core,
then we expect $B_r\approx B_\theta$, and $J_\phi\approx
(c/4\pi)(B_\theta/R)$ giving a Hall timescale a factor $R/H$
longer. This is directly analogous to the sensitivity of the lowest
order Ohmic decay mode to the boundary condition at the base that we
found in \S \ref{sec:ohmicevolution}.

\subsection{Time-Independent Solutions}

Not all magnetic field configurations evolve under the Hall
effect. The existence of time-independent solutions may be important
for two reasons. First, it implies that dissipation by the Hall
effect is not necessarily able to cause the field to decrease to
arbitrarily small values (assuming that the role of the Hall effect is to drive dissipative processes on small scales). Second, time-independent solutions may act as a background on which Hall waves
propagate. The most obvious time-independent solution is a force free
field with $\vec{J}\vcross\vec{B}=0$, or $\vec{J}=\kappa \vec{B}$,
where $\kappa$ is constant along a given field line (e.g. Mestel
1999). The field profile in this case may or may not reflect
lengthscales associated with the crust, depending on the initial
relaxation of the field when the neutron star is born.

An intriguing time-independent solution is suggested by equation
(\ref{eq:poloidalevol}). For an axisymmetric dipole field, the
toroidal current is $J_\phi\propto\sin\theta$
(e.g.~Fig.~\ref{fig:twist}a), implying that $\Omega_e\propto
v_e/\sin\theta$ is constant, or the electrons rotate rigidly on
spherical shells. Equation (\ref{eq:poloidalevol}) then becomes
$\partial B_\phi/\partial t=r\sin\theta B_r(\partial\Omega_e/\partial
r)$. A simple time-independent solution is therefore
$\Omega_e=$constant, or {\it rigid rotation of the electrons}! If the
electrons rigidly rotate, no shearing of the poloidal field occurs,
and therefore no evolution due to the Hall effect. This solution
implies $J_\phi\propto n_e$, and therefore large currents running at
the base compared to the top of the crust. If this current is
sufficiently large compared to currents in the core, then deep in the
crust the field is $B_\theta\simeq -(4\pi
J_{\phi,b}H_e/c)(n_e/n_{e,b})$, $B_r\simeq 2B_\theta(H_e/r)$, where
$J_{\phi,b}$ and $n_{e,b}$ are the current and electron density at the
base. This solution is not force free, but implies an increasing
magnetic force with depth; the magnetic stresses are poloidal however,
and hence may be supported in the fluid above and below the crust.

It is interesting to speculate whether further time-independent
solutions exist. Equation (\ref{eq:poloidalevol})
suggests, analogous to Ferraro's law of isorotation (Ferraro 1937; Mestel 1999), further
steady-states in which the electrons rigidly rotate along each
poloidal field line. However, the difference between the Hall effect
and the MHD case is that the electron velocity depends directly on the
magnetic field. The poloidal dipole field is special because the
currents are such that no differential rotation of the electrons
occurs on spherical shells. This is not true in general, and therefore
the dipole case may be the only steady-state solution of this
kind. A toroidal field in plane-parallel geometry $\vec{B}=B(x,z)\hat e_y$ (where we assume the field is "axisymmetric", $\partial/\partial y=0$), and with constant $n_e$, does not evolve due to the Hall term, since $\curl(\vec{J}\vcross\vec{B})=0$. However, Vainshtein et al.~(2000) noted that a gradient in $n_e$ gives rise to compression of the field (second term on the right hand side of eq.~[\ref{eq:toroidalevol}]), and concluded that the field is not steady-state in general. This is likely also true once the extra terms arising from spherical effects (eq.~[\ref{eq:toroidalevol}]) are included, unless the angular distribution of the extra terms can be arranged to match (and cancel) the compression term.

The time independent solution is stable to small perturbations.
In order to see this, let
$\vec{B}$ and $\vec{J}$ be the magnetic field and current density of the
time independent state, and $\delta\vec{B}$ and $\delta\vec{J}$ be
the corresponding perturbations of this state. Linearizing eq. (\ref{eq:inducthall}) with respect to the perturbations gives
\begin{equation}\label{eq:lininducthall}
\frac{\partial\delta\vec{B}}{\partial t}=-\grad\vcross\left(\frac{\vec{J}
\vcross\delta\vec{B}+\delta\vec{J}\vcross\vec{B}}{en_e}\right).
\end{equation}
Taking the scalar product of eq. (\ref{eq:lininducthall}) with $\delta\vec{B}/4\pi$ and integrating over the volume of the crust gives an equation for
the magnetic energy in the perturbation
\begin{equation}\label{eq:energy1}
\frac{\partial}{\partial t}\int\frac{\delta B^2}{8\pi}dV=\int\frac{\vec{J}}{en_ec}\cdot\left(\delta\vec{J}
\vcross\delta\vec{B}\right)dV.
\end{equation}
In deriving eq. (\ref{eq:energy1}),
we have integrated the right hand side once by parts, used Ampere's
Law, and assumed the perturbations vanish on the boundary. Next, we rewrite  
$\delta\vec{J}\vcross\delta\vec{B}/c$ as the divergence of the Maxwell tensor
\begin{equation}\label{eq:deltajdeltab}
\frac{\delta\vec{J}\vcross\delta\vec{B}}{c}=\grad\cdot\vec{T},
\end{equation}
where
\begin{equation}\label{eq:T}
\vec{T}\equiv\frac{\delta\vec{B}\delta\vec{B}}{4\pi}-\vec{I}\frac{\delta B^2}{8\pi}.
\end{equation}
We need only the $T_{i\phi}$ components of eq. (\ref{eq:T}) because $\vec{J}$
has only a $\phi$ component;
$\vec{J}/en_e =-\hat\phi\Omega_er\sin{\theta}$, with
$\Omega_e$ a constant. Writing out the integrand on the RHS of eq. (\ref{eq:energy1}), using eqs. (\ref{eq:deltajdeltab}) and (\ref{eq:T}), and applying
Gauss's Theorem, we find that the integral vanishes identically. This shows
that in the absence of any dissipation, small perturbations to the steady
state neither grow nor decay. Although the realizability of this steady state
is open to question, the demonstration that it is stable certainly bolsters
its case for relevance. 

Lastly, we mention that this background solution will decay with time at a rate
given roughly by the Ohmic time at the base of the crust (section \ref{sec:ohmictime}).


\subsection{Hall waves}
\label{sec:hall_uniform}

To make further progress, we consider the evolution of perturbations
to a background magnetic field. GR showed that such perturbations
propagate as circularly-polarized {\em Hall waves}. For constant
electron density and a uniform magnetic field, the induction equation
describing plane-wave perturbations $\delta B\propto
\exp(i\vec{k}\cdot\vec{x})$ is
\begin{equation}\label{eq:hall_rotate}
{\partial\delta\vec{B}\over\partial
t}={c\vec{k}\cdot\vec{B}\over 4\pi
n_ee}\left(\vec{k}\vcross\delta\vec{B}\right),
\end{equation}
which may be derived by perturbing equation (\ref{eq:inducthall}) (see
also Thompson \& Duncan 1996). The dispersion relation is
\begin{equation}\label{eq:hall_dispersion}
\omega={ck\left|\vec{k}\cdot\vec{B}\right|\over 4\pi n_e e}= \nu k\left|\vec{k}\cdot\hat e_B\right|,
\end{equation}
where we define $\nu\equiv cB/4\pi n_e e = 
\omega_{ci}\delta_i^2\propto B/n_e$. GR discussed
the suggestion by Jones (1988) that Hall waves could transport energy
to lower densities in the crust, where Ohmic dissipation is more
efficient. They pointed out that as the wave moves to lower $n_e$, the
vertical wavelength must increase (for fixed $\omega$; see
eq.~[\ref{eq:hall_dispersion}]), implying a reflection or turning point for the
wave when $k_zH\sim 1$. In this section, we calculate the properties
of Hall waves in a realistic crust model. We determine the location of
the turning point as a function of wavelength. In addition, we
calculate the strain induced by the Hall wave in the crust, and discuss unstable modes in the presence of a background electron velocity gradient.

\subsubsection{Force balance in the solid}

The Hall wave in the crust relies on the fact that the solid can
adjust to accommodate the $\vec{J}\vcross\vec{B}$ forces that arise as
the wave propagates (see also Franco et.al. (2000) for the case of magnetic stresses
induced by spin down). To see this, consider perturbations
$\propto\exp(-i\omega t+i\vec{k}\cdot\vec{x})$ of a uniform background
field with constant $n_e$ as before, but now include the shear stress
(with constant shear modulus $\mu$) and the ion inertia. Since we
assume constant $n_e$, the orientation of the field is not
important. The induction equation (\ref{eq:hall_rotate}) has an extra
term $\curl\left(\delta\vec{v}\vcross\vec{B}\right)$, giving
\begin{equation}
-i\omega\delta\vec{B}={c\vec{k}\cdot\vec{B}\over4\pi n_ee}
\left(\vec{k}\vcross\delta\vec{B}\right)+
\vec{k}\vcross\left(\omega\vec{\xi}\vcross\vec{B}\right),
\end{equation}
where we write the velocity in terms of the fluid displacement as
$\delta\vec{v}=i\omega\vec{\xi}$. The momentum equation is
\begin{equation}
-\rho\omega^2\vec{\xi}={\delta\vec{J}\vcross\vec{B}\over c}-\mu k^2\vec{\xi},
\end{equation}
with the terms representing the ion inertia, magnetic force, and shear
stress. Solving for $\omega$, we find the dispersion relation
\begin{equation}\label{eq:hall_general_dispersion}
\omega^2={\omega_{\rm Hall}^2\left(\omega^2-\omega_s^2\right)\over
\left(\omega^2-\omega_s^2-\omega_A^2\right)},
\end{equation}
where $\omega_s^2=k^2\mu/\rho$ and
$\omega_A^2=(\vec{k}\cdot\vec{B})^2/4\pi\rho$ define the shear and
Alfven wave frequencies, and $\omega_{\rm
Hall}=ck(\vec{k}\cdot\vec{B})/4\pi n_ee$ is the Hall wave frequency.

In the solid crust, which has a finite shear modulus $\omega_s^2>0$,
there is a low frequency ($\omega^2\ll \omega_A^2,\omega_s^2$)
solution to equation (\ref{eq:hall_general_dispersion}),
\begin{equation}
\omega={\omega_{\rm Hall}\over 1+B^2/4\pi\mu}.
\end{equation}
For $B^2/4\pi \ll \mu$, we recover the usual Hall wave: the induction
and momentum equations are uncoupled. For $B^2/4\pi \gg \mu$, the emf
induced by the fluid displacements is large enough to counteract the
Hall emf, reducing the Hall wave frequency. For typical pulsar fields,
this is an important effect only at the very top of the crust; for
magnetar-strength fields, it is an important effect throughout much of
the crust. The shear modulus is $\mu\approx 0.1n_i(Ze)^2/a$
(Strohmayer et al.~1991), where $a=(3/4\pi n_i)^{1/3}$ is the interion
spacing\footnote{A convenient way to write this when relativistic
degenerate electrons provide the pressure is $\mu/P_e\approx 10^{-2}\
(Z/26)^{2/3}$.}, giving $B^2/4\pi\mu=5\times 10^{-3} B_{12}^2
(\rho_9Y_e)^{-4/3}(Z/26)^{-2/3}$. In this paper, we assume $B^2\ll4\pi\mu$, so that the induction and momentum equations are uncoupled in the crust. 

In the fluid above or below the crust, the shear modulus is
zero, so that the $\vec{J}\vcross\vec{B}$ force must be balanced by
the ion inertia, giving a natural wave frequency
$\omega\approx\omega_A\gg\omega_{\rm Hall}$. One might expect that the Hall waves would be almost completely reflected at the crust/fluid boundaries because of this extremely large mismatch in wave frequency. 
However, deriving the behavior of the strain $\theta=d\xi/dz$ near the boundaries requires a careful treatment of the boundary conditions. In this paper, we calculate the response of the crust for the simplest case of a vertically propagating wave on a vertical background field. We discuss the behavior of the wave at the boundary in the Appendix.
The key point is that the Hall term enforces continuity of the magnetic
perturbation across the boundary. Therefore, the magnetic stress
cannot abruptly change to accommodate the loss of shear stress on
moving into the fluid. This implies that $\mu\theta\approx B\delta
B/4\pi\approx 0$ at the fluid/solid interface, which gives complete reflection. 

\subsubsection{Vertically-propagating waves including the density gradient}

We now include the electron density gradient, and study
vertically-propagating Hall waves in a uniform vertical background
field. In terms of helicity
eigenstates $\delta B_{\pm}(z)=\delta B_x(z)\pm i\delta B_y(z)$, the
perturbed induction equation is
\begin{equation}\label{eq:vertequation}
{d\over dz}\left(\nu{d\delta B_{\pm}\over dz}\right)\pm\omega\delta
B_{\pm}=0.
\end{equation}
Since $\divr\vec{\delta B}=0$, $\delta B_z=\ $constant for these
modes.

\begin{figure}
\epsscale{1.0}\plotone{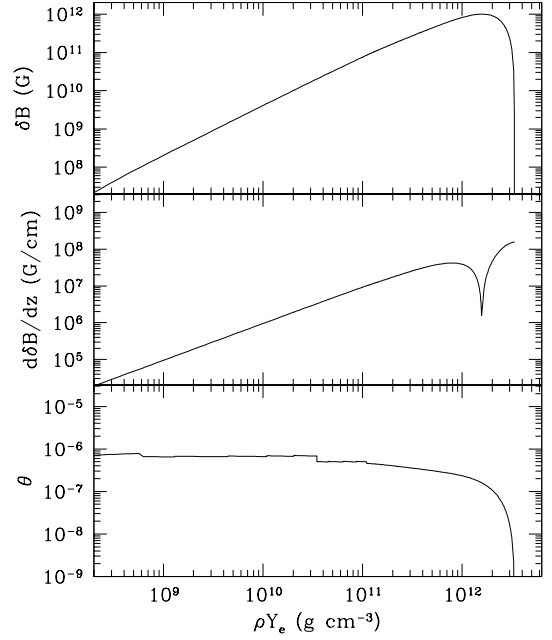}
\caption{Hall wave with no nodes in a constant vertical background
field of $B=10^{12}\ {\rm G}$. This wave has a period $P_1=4.4\ \times
10^6\ {\rm yr}\ B_{12}^{-1}$. We normalize the wave so that it has a
maximum amplitude $\delta B/B=1$. At the top of the crust (not shown),
$\theta$ drops rapidly to zero (see text).}
\label{fig:hallmode1}
\end{figure}

\begin{figure}
\epsscale{1.0}\plotone{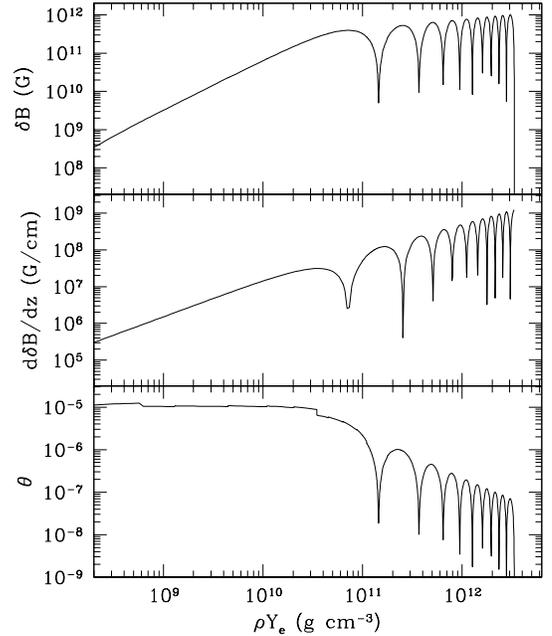}
\caption{As Figure \ref{fig:hallmode1}, but for 10 nodes. This wave
has a period $P_{10}=5.7\ \times 10^4\ {\rm yr}\ B_{12}^{-1}$.}
\label{fig:hallmode10}
\end{figure}

\begin{figure}
\epsscale{1.0}\plotone{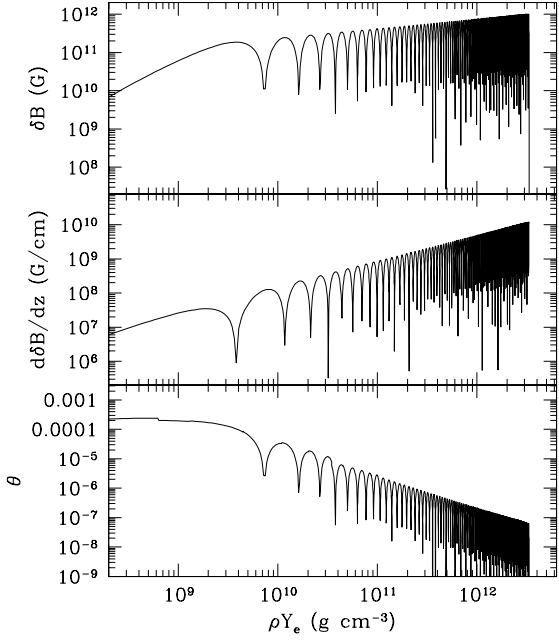}
\caption{As Figure \ref{fig:hallmode1}, but for 100 nodes. This wave
has a period $P_{100}=580\ {\rm yr}\ B_{12}^{-1}$.}
\label{fig:hallmode100}
\end{figure}

We integrate equation (\ref{eq:vertequation}) through the crust, using
a shooting method to find the eigenvalues $\omega$ for which $\delta
B=0$ at each fluid/solid interface. Figures
\ref{fig:hallmode1}--\ref{fig:hallmode100} show the eigenmodes with
$n=1,10,$ and $100$ nodes, normalized so that the maximum perturbation
is $\delta B=10^{12}\ {\rm G}$.

The properties of the eigenmodes can be simply understood with a WKB
analysis. A good fit to the electron density profile in the crust is
$n_e(z)\simeq n_{e,b} (z/4H_b)^4$, with $n_{e,b}=3\ \times 10^{36}\
{\rm cm^{-3}}$ and $H_b\simeq 250\ {\rm m}$ the electron density and
scale height at the base of the crust. Using this to quantize the
number of nodes $n$ over the crust, $\int k_z dz=n\pi$ gives the wave
period
\begin{equation}
P_{\rm Hall}\simeq {t_{\rm Hall,b}\over n^2}\simeq {10^7\ {\rm yrs}\over
B_{12}n^2}.
\end{equation}
The WKB solution gives $\delta B\propto k^{1/2}\propto
n_e^{1/4}\propto z$, in good agreement with the numerical
solutions. The wave reaches a turning point when $\int k_z dz\sim 1$,
or at a depth $z_{\rm turn}\approx 3 H_b/n^{1/3}$. In terms of
density, the turning point is
\begin{equation}
\rho_{\rm turn}Y_e\approx {2\times 10^{12}\ {\rm g\ cm^{-3}}\over
n^{4/3}},
\end{equation}
which requires $n\approx 300$ for $\rho_{\rm turn}\approx 10^9\ {\rm
g\ cm^{-3}}$, or $n\approx 3\times 10^5$ for $\rho_{\rm turn}\approx
10^5\ {\rm g\ cm^{-3}}$. Above the turning point, $\delta B$ decreases
to match the boundary condition $\delta B=0$. Since $\nu\propto
z^{-4}$, equation (\ref{eq:vertequation}) gives a steep dropoff
$\delta B\propto z^5$, in agreement with our numerical results.

The Ohmic dissipation rate for the mode is
\be
\gamma & = & \left. \int dV {\left|\delta\vec{J}\right|^2\over\sigma}
\right/  \int dV {\left|\delta\vec{B}\right|^2\over 4\pi}
\nonumber \\
& \simeq & 
\left. \int_{z_{tp}}^{z_b} \frac{dz}{v_{g,z}}\frac{c^2k_z^2}{4\pi \sigma}
 \right/ \int_{z_{tp}}^{z_b} \frac{dz}{v_{g,z}},
\ee
where $v_{g,z}$ is the vertical group velocity, 
and in the second expression we have used constant wave energy flux 
in the propagation cavity, and the integral extends between the upper turning
point $z_{tp}$ and the base of the crust $z_b$.
This expression is simply interpreted as the local damping rate weighted by the time
spent at that depth. 
For $k_z \gg k_\perp$, $v_{g,z} \propto k_z^{-1}$
so the ratio $zk_z^3/\rho^{1/3}$ determines the depth which dominates Ohmic decay
(for impurity scattering). If $k_z\propto n_e^{1/2}$, the factor $n_e^{3/2}/\rho^{1/3}$
is dominated by the largest densities. We expect
that up to an order unity prefactor, the Ohmic decay time for the mode
is set by $\Omega\tau$ at the base of the crust, $\gamma^{-1}\approx
P_{\rm Hall}(\Omega\tau)_b$, where $(\Omega\tau)_b\approx B_{12}/Q$
when impurity scattering dominates (eq.~[\ref{eq:omegatau1}]).

\subsubsection{The elastic response of the crust to the wave}

The strain in the crust may be calculated by balancing the elastic and
magnetic stresses, since ion inertia plays no role in the force
balance for the low frequency Hall waves. This gives
\begin{equation}
\theta={d\xi\over dz}= -{B_z\over 4\pi\mu} \delta B.  \label{eq:strain}
\end{equation}
We show the strain as a function of depth in the lower panels of Figures
\ref{fig:hallmode1}--\ref{fig:hallmode100}. We write the shear modulus
(Strohmayer et al.~1991) as $\mu\propto n_e^{4/3}\approx 10^{30}\
{\rm erg\ cm^{-3}}\ (z/z_b)^{16/3}$, giving $\theta\propto z^{-13/3}$
below the turning point, and $\theta\propto z^{-1/3}$ above the
turning point. The maximum strain occurs at or above the turning
point. Using the WKB scaling, we find that at the turning point,
\begin{equation}\label{eq:theta_max}
\theta_{\rm turn}=3\times 10^{-7}\ B_{12}^2 n^{13/9} {\delta
B_b\over B},
\end{equation}
where $\delta B_b$ is the amplitude of the mode at the base.

\subsubsection{More general cases of Hall waves}

For a horizontal background field, or a non-vertically propagating wave on a vertical background field, the calculation of the elastic response of the crust is more complicated, and we leave this for future work. Here, we give the WKB scalings for waves on a uniform horizontal field $\vec{B}=B\hat e_x$, and set $k_y=0$ for simplicity, $\delta B=\delta \vec{B}(z)\exp(ik_xx-i\omega t)$. The perturbation equation is 
\begin{equation}\label{eq:hall_horizontal}
{d^2\delta B_z\over dz^2}+k_x^2\delta B_z\left[{\omega^2\over
\nu^2k_x^4}-1\right]=0,
\end{equation}
giving the WKB dispersion relation $\omega=\pm\nu k_x(k_x^2+k_z^2)^{1/2}$. For $k_z\gg k_x$, this gives $P_{\rm Hall}\approx t_{\rm
Hall}/(n k_xH_b)$. The location of the turning point is $z_\mathrm{turn}\approx 3H_b/n^{1/5}$, or $\rho_\mathrm{turn}Y_e\approx 2\times 10^{12}\ \mathrm{g\ cm^{-3}}/n^{4/5}$. The turning point for the modes is deeper in than the vertically-propagating case; we require $n\approx 10^4$ for $\rho_\mathrm{turn}\approx 10^9\ \mathrm{g\ cm^{-3}}$.


\subsubsection{Non-zero background current: waves and unstable modes}

Before ending this section, we discuss the case of a horizontal field in more detail.
In several recent papers, Rheinhardt, Geppert, and collaborators have
discussed a ``Hall drift instability'' (HDI) driven by curvature in the
background field (Rheinhardt \& Geppert 2002, hereafter RG02; Geppert
\& Rheinhardt 2002; Rheinhardt, Konenkov, \& Geppert 2003; Geppert,
Rheinhardt, \& Gil 2003). In this section, we first discuss a simple
example of the instability to elucidate the basic physics. We then
discuss the mode spectrum, and the relevance of the instability for
small scale Hall waves.

Following RG02, we study the stability of a plane-parallel horizontal
field that depends only on height $z$, $\vec{B}=B(z)\hat{e}_x$, with
corresponding current density $\vec{J}=(c/4\pi)(dB/dz)\hat{e}_y$. However, for simplicity and to highlight the essential physics, we consider the Hall terms only. The perturbations are of the form $\delta \vec{B}(z)\exp(ik_xx+\sigma t)$ (we set
$k_y=0$ since RG02 find that these are the fastest growing
modes). Since the background field does not evolve due to the Hall
term, the perturbed induction equation is
$\sigma\delta\vec{B}=\curl(\delta\vec{v_e}\vcross\vec{B})+\curl(\vec{v_e}\vcross\delta\vec{B})$,
or
\begin{equation}\label{eq:hdi_induct}
\left(\sigma+\vec{v_e}\vdot\grad\right)\
\delta\vec{B}=\left(\vec{B}\vdot\grad\right)\delta\vec{v_e}
+\left(\delta\vec{B}\vdot\grad\right)\vec{v_e}
-\left(\delta\vec{v_e}\vdot\grad\right)\vec{B},
\end{equation}
where $\vec{v_e}=-\vec{J}/n_ee$ is the electron velocity, and
$\delta\vec{J}=(c/4\pi)(\curl\delta\vec{B})$. 

Writing an equation for $\delta B_z$, we find the generalization of equation (\ref{eq:hall_horizontal}) for this case is,
\begin{equation}\label{eq:hdi_bz}
{\partial^2\delta B_z\over\partial z^2}=k_x^2\delta B_z
\left[1+{dv_e/dz\over\nu k_x^2}+{\sigma^2\over (\nu k_x^2)^2}
\right].
\end{equation}
The eigenvalue $\sigma^2$ is purely real\footnote{To show this,
multiply equation (\ref{eq:hdi_bz}) by the complex conjugate $\delta
B_z^\star$, and subtract the complex conjugate of equation
(\ref{eq:hdi_bz}) multiplied by $\delta B_z$. Integrating over volume,
and assuming the boundary conditions are such that the boundary term
vanishes, gives $(\sigma^2-\sigma^{2\star})\int dV \left|\delta
B_z\right|^2/\nu^2k_x^2=0$.}: the solution is either a pure
growing/decaying mode, or oscillatory. Making a WKB approximation,
$\partial/\partial z\rightarrow ik_z$, we find (writing
$k^2=k_x^2+k_z^2$),
\begin{equation}\label{eq:hdi_disp}
\sigma^2=-(\nu k k_x)^2 - \nu k_x^2 {dv_e\over dz}.
\end{equation}
The first term is the usual Hall waves, $\sigma=\pm i \nu k k_x$; the
second term depends on the electron velocity gradient, and leads to
instability ($\sigma^2>0$) if the electron velocity gradient is
sufficiently negative, and for sufficiently long wavelengths,
\begin{equation}\label{eq:hdi_wkb_criterion}
k^2 < -{1\over\nu}{dv_e\over dz}.
\end{equation}
A more general integral criterion can be derived by multiplying
equation (\ref{eq:hdi_bz}) by $\delta B_z$ and integrating over the
volume,
\begin{eqnarray}
\sigma^2\int dV{\left|\delta B_z\right|^2\over \nu^2k_x^2}= -\int
dV\left(\left|\delta B_z^\prime\right|^2+k_x^2\left|\delta
B_z\right|^2\right)\nonumber\\
-\int dV\left|\delta B_z\right|^2{dv_e/dz\over
  \nu},
\end{eqnarray}
showing that $dv_e/dz$ must be sufficiently negative somewhere in the
volume to lead to instability ($\sigma^2>0$). We have numerically
integrated equation (\ref{eq:hdi_bz}) for the simple case of constant
$n_e$ and a quadratic dependence of $B$ on depth, confirming the
scalings of equations (\ref{eq:hdi_disp}) and
(\ref{eq:hdi_wkb_criterion}).

The basic physics of the instability can be understood by stepping
through the evolution of an initial vertical perturbation $\delta
B_z$, and recalling that the field is advected by the electron flow. First
consider short wavelengths where the $dv_e/dz$ term is
unimportant. The current perturbation $\delta J_y=-ik_x\delta
B_zc/4\pi$ corresponds to a sheared electron flow, which gives rise to
a perpendicular field component $\sigma \delta B_y=-ik_xB\delta
v_{e,y}$ through the first term on the right hand side of equation
(\ref{eq:hdi_induct}). Associated with $\delta B_y$ is a current
$\delta J_z=ik_x\delta B_yc/4\pi$, which in turn acts back on $\delta
B_z$, $\sigma\delta B_z=-ik_xB\delta v_{e,z}$, thereby driving the
Hall wave. When the $dv_e/dz$ term dominates, the first stage of this
evolution is different. Shearing of the initial perturbation by the
background electron flow generates $\delta B_y$ in the first step,
$\sigma\delta B_y=-\delta B_z(dv_e/dz)$, transferring energy from the
background field into the perturbation. Equation (\ref{eq:hdi_disp})
shows the competition between the internal and external shearing,
measured by the ratio of Hall wave frequency to the shear frequency
$dv_e/dz$.

The criterion for instability we derive above generalizes that of RG02
to include the density variation with depth (see also Rheinhardt et
al.~2003), and clarifies the nature of the instability: a background shear in the electron velocity drives growth of long wavelength perturbations whose characteristic Hall frequencies are less than $dv_e/dz$. Short wavelength Hall waves, with frequencies greater than $dv_e/dz$, are unaffected. This differs from the conclusions of RG02, who argue that HDI may be responsible for the growth of small-scale perturbations reported in some numerical simulations (e.g.~Shalybkov \& Urpin 1997; Hollerbach \& R\"udiger 2002). Perhaps numerical instabilities are rather responsible for the excess small scale power\footnote{We found such an instability in trying to develop a 3D code to simulate the Hall effect. It is straightforward to see this in 1D in Cartesian coordinates. Using second-order centered differences in space and explicit time-stepping (first order, or, using Runga-Kutta, second order accurate), one can analytically show that {\it all} wavelengths are unstable! The growth rate of the instability for wavenumber $k$ on a grid of spacing $\Delta x$ scales as $\propto (k\Delta x)^8$, implying large lengthscale waves grow only after many box crossings, but grid-scale waves have a growth rate of order the time step. Including Ohmic diffusion only cures this instability if a prohibitively short time step is used. In 1D, the instability is easily fixed by using implicit time-stepping, but this cannot be simply extended to 3D using operator splitting.}.

Is the HDI relevant for the evolution of field in the crust? If the picture of a turbulent Hall cascade is correct, the instability probably does not change the long-term evolution of the field, since intermediate scales will "fill in" as the cascade develops. If the background field varies on a lengthscale $H$, the most unstable modes are those with the largest $k_x$ compatible with equation (\ref{eq:hdi_wkb_criterion}), implying $k_x\sim 1/H$, or corresponding to spherical harmonic $l$ of tens. Rheinhardt et al.~(2003) then argue that an initially dipolar field will develop small latitudinal scale structure; however, this conclusion is sensitive to the initial distribution of currents in the crust. An additional point is that changes in the crustal field may not propagate to the surface because of the very efficient reflection at the solid/fluid interface (see discussion in the Appendix). Therefore predictions of complex surface fields produced by HDI (Geppert et al.~2003) may be premature.


\subsection{Nonlinear Coupling of Oscillation Modes}
\label{sec:nonlinear}

We now discuss the implications of our results for the turbulent Hall cascade suggested by GR. They estimated the non-linear transfer time between different modes to
be longer than the mode period, and proposed that the cascade would be
a weak cascade. In this case, the expected scaling of magnetic energy
is $B^2(k)\propto k^{-2}$, with a cutoff due to Ohmic dissipation at
a scale $\lambda\sim L/\Omega\tau$, where $L$ is the scale on which
the turbulence is being driven. However, the simulations of Biskamp et
al.~(1999) show a different scaling, $B^2(k)\propto k^{-7/3}$
appropriate for a strong cascade (to derive this scaling, follow the
argument of GR \S 4, but choosing their eq.~[47] rather than
eq.~[48]). In this case, the Ohmic cutoff is for $\lambda\sim
L/(\Omega\tau)^{3/2}$. The difference arises most likely because of
GR's assumption of isotropy: a strong cascade implies mode
periods longer than the non-linear transfer time, which suggests a cascade perpendicular to the magnetic field direction, for which the Hall wave periods are long. Unfortunately, Biskamp and coauthors did not present an analysis of anisotropy in their numerical results. Such a study would be extremely interesting.

How many modes are involved in the cascade? If the energy transfer is limited by Ohmic dissipation, then for a strong cascade we find
\begin{equation}\label{eq:nmax}
n_\mathrm{max}\approx (\Omega\tau)^{3/2}\approx 3\times 10^4\
\left({B_{12}\over Q_{-3}}\right)^{3/2}
\end{equation}
where we use the result for impurity scattering for $\Omega\tau$. The
turning point for the mode with $n=n_\mathrm{max}$ (assuming vertical propagation and a vertical field) is at $\rho_\mathrm{max}\approx 10^6\ {\rm g\ cm^{-3}}\ (Q_{-3}/B_{12})^2$.

In \S \ref{sec:hall_uniform}, we calculated the maximum strain
$\theta_{\rm max}$ for a Hall wave with given $n$ and amplitude at the
base $\delta B_b$. As $n$ increases, $\theta_{\rm max}$ increases,
implying that the crust will yield for large enough $n$. However, this
increase with $n$ is offset by the decreasing amplitude $\delta B_b$
with $n$ if there is a turbulent cascade: for a strong (weak) cascade,
$\delta B\propto n^{-7/6}$ ($\propto n^{-1}$). The value of strain at
which the crust will yield $\theta_{\rm yield}$ is uncertain
(e.g.~Ruderman 1976), but probably lies in the range $\theta_{\rm
yield}\sim 10^{-5}$--$10^{-2}$. For the weak cascade, $\theta_{\rm
max}$ decreases more slowly with $n$. Using equation
(\ref{eq:theta_max}) for vertically-propagating waves, we find for
this case that the crust will yield for modes with
\begin{equation}\label{eq:ncrack}
n\gtrsim 6\times 10^5\ \left({\theta_{\rm yield}\over
10^{-4}}\right)^{\beta}B_{12}^{-2\beta}\left({\delta B_{b,0}\over
B}\right)^{-\beta},
\end{equation}
where $\delta B_{b,0}$ is the amplitude of the mode at the largest
lengthscale, and $\beta=9/4$. For a strong cascade, the prefactor is
$2\times 10^9$ (corresponding to $\Omega\tau\approx 10^6$) and
$\beta=3.2$. Given our estimate for $n_\mathrm{max}$ (eq.~[\ref{eq:nmax}]), we therefore do not expect the cascade to be limited by crust yielding, but rather Ohmic dissipation. However, we stress again that we have not yet calculated the crust response for arbitrary wave propagation and background field directions. Therefore, equation (\ref{eq:ncrack}) should be taken as an initial estimate only.


\section{Summary and Conclusions}

In this paper, we have addressed various aspects of magnetic field evolution in neutron star crusts. Our aim has been to place previous studies of the physics of the Hall effect in context: first, by addressing when and for which neutron stars the Hall effect is important relative to Ohmic decay; second, by calculating the allowed "initial conditions" by following the Ohmic decay prior to dominance of the Hall effect; third, by discussing for the first time the properties of Hall waves in a realistic crust model, and the implications for the Hall cascade. 

We now summarise our main results and conclusions:

(i) The relative importance of Ohmic decay and the Hall effect is summarized in Figure \ref{fig:bt} which shows the $\Omega\tau=1$ contour in the $B$-$T$ plane. In isolated neutron stars with $B\lesssim 10^{13}\ \mathrm{G}$, the dominance of the Hall effect depends sensitively on the impurity parameter $Q$: recent calculations by Jones (2001) indicate $Q\gtrsim 1$ for densities greater than neutron drip (see also de Blasio 2000 for densities lower than neutron drip), in which case the Hall effect never dominates the evolution of crustal currents, considerably simplifying the evolution. If $Q\sim 10^{-3}$ however, as in the original Flowers \& Ruderman (1977) estimate, Ohmic decay dominates the Hall effect only for a time  $t_{\rm switch}\approx 2\times 10^4\ {\rm years}/B_{12}^3$, after which the conductivity is set by impurity scattering, giving $\Omega\tau\approx B_{12}/Q$. In accreting neutron stars with $B\lesssim 10^{13}\ \mathrm{G}$, Ohmic decay always dominates, since the crust is heated ($T\gtrsim 10^8\ {\rm K}$, e.g.~Brown 2000) and may also be impure $Q\gtrsim 1$ (Schatz et al.~1999).

(ii) Figure \ref{fig:ohmictime_vs_depth} summarizes the evolution of currents in the crust while Ohmic decay dominates ($t<t_{\rm switch}$). The electrical conductivity at these early times is set by phonon scattering. This figure relates previous work on either Ohmic decay eigenmodes, or on self-similar solutions, in a simple way. For a particular time $t$, Figure \ref{fig:ohmictime_vs_depth} gives the density $\rho_c$ down to which the currents have significantly decayed. For $\rho>\rho_c$, no significant evolution of the currents has occurred. For $\rho<\rho_c$, the current distribution is self-similar, being dominated by the Ohmic decay mode with turning point at $\rho=\rho_c$. Because $t_{\rm switch}\propto 1/B^3$, the allowed distribution of currents when the Hall effect begins to dominate is very sensitive to $B$. For $B\approx 10^{13}\ {\rm G}$, $\rho_c$ lies at or above the top of the crust; for $B\approx 10^{11}\ {\rm G}$, $\rho_c$ lies at the base of the crust. At late times, when $\rho_c$ is greater than the initial location of the currents, the current distribution evolves in a self-similar manner and the surface field decays as a power law (eq.[\ref{eq:fielddecay}]).

(iii) The basic physics of the Hall effect is that the field in the crust is advected by the electron flow. This is summarized in Figure \ref{fig:shear}, which considers a variant of the usual laboratory demonstration of the Hall effect. Figure \ref{fig:twist} shows the evolution of an initial poloidal dipole field, in which the toroidal currents initially "twist" the field; the resulting poloidal currents then generate a quadrupole poloidal field via a "pinch". These considerations led us to a simple time-independent solution: a poloidal dipole field, in which the electrons rotate rigidly, does not evolve due to the Hall effect.

(iv) We discussed the properties of Hall waves in the crust. The simplest example is a background magnetic field which is uniform and vertical (for example, emerging from the core). The vertical wavelength of the wave grows towards the top of the crust (as pointed out by GR), and we showed that the turning point location is $\rho_{\rm turn}\approx 10^{12}\ {\rm g\ cm^{-3}}/n^{4/3}$. Below the turning point, the wave is able to propagate and its properties are well-described by WKB scalings. Above the turning point, the wave evanesces towards the top of the crust.

(v) A horizontal field (e.g.~as may be produced if the currents supporting the field are predominantly in the crust) has similar wave-like solutions at short wavelengths. However, if there is a background current (the background $B$ varies with height), long wavelength perturbations become unstable. The physics of this instability is that the shear in the background electron velocity drives unstable growth, overcoming the internal shear of the wave which would otherwise drive an oscillation. Therefore we agree with the recent work of Geppert and collaborators that there is an instability, and we have clarified its basic physics. However, we argued that the relevance of this instability for the field evolution in the crust, and particularly for the observable surface field, is not clear.

(vi) We have made a first attempt to address the question of whether Hall waves strain the crust beyond its yield point. We find that the maximum strain is at or above the turning point for the mode (eq.~[\ref{eq:theta_max}]). However, our calculation is limited to the special case of vertically-propagating waves on a vertical background field; we leave further calculations of the response of the crust to non-vertical background fields and wavevectors for future work. In particular, the force balance is important to consider for magnetar strength fields, for which $B^2>4\pi\mu$, directly coupling the induction and momentum equations in the crust (see the recent discussion in Arras et al.~2004).

(vii) Finally, we briefly discussed the non-linear evolution. Recent numerical work on Whistler turbulence (Biskamp et al.~1999) has demonstrated GR's conjecture that the Hall effect would create a "cascade" of energy to small scales. However, the scalings of the cascade imply strong, anisotropic turbulence rather than weak, isotropic turbulence as assumed by GR. The Hall wave eigenfunctions calculated in \S 4 (again, for vertically propagating waves and a vertical field) show that crust breaking is probably not important for standard radio pulsar strength fields.

\acknowledgements 
We thank Omer Blaes, Phil Chang, Lars Bildsten, Roger Romani, Chris
Thompson, Steve Thorsett, and Dmitry Uzdensky for useful discussions,
and Peter Goldreich for stressing the simple physics that the magnetic
field moves with the electrons. AC acknowledges support from NASA
through Hubble Fellowship grant HF-01138 awarded by the Space
Telescope Science Institute, which is operated by the Association of
Universities for Research in Astronomy, Inc., for NASA, under contract
NAS 5-26555. PA acknowledges support from the Canadian Institute for
Theoretical Astrophysics. PA is an NSF Astronomy and Astrophysics
Postdoctoral Fellow. This material is based upon work supported by the
National Science Foundation under Grant No. 0201636 to the University of
California at Santa Barbara and Grants  AST-0328821 and PHY-0215581 to the 
University of
Wisconsin at Madison.

\appendix

\section{ Boundary conditions for Hall waves }

In this Appendix we discuss the boundary conditions for Hall waves in the
crust. First we argue that the disparity in propagation speeds between
solid and liquid implies near total reflection of wave energy at the
boundaries. We present the results of simple toy models to quantify the
transmission coefficient. We then derive the boundary conditions. Lastly
we specialize the boundary conditions for vacuum solutions in the liquid.

For elastic response of the crust, i.e. neglecting irreversible
processes such as plastic flow, we naively expect Hall wave energy to
be almost perfectly trapped in the crust. For a wavelength of order the
size of the crust, the Hall wave propagation speed is vastly smaller
than the Alfven speed, implying a large impedence mismatch between the
solid crust, liquid core and liquid ocean.  

To get an analytic handle on the size of the reflection and transmission
coefficients, consider a region with uniform density, constant vertical
background field, crust with constant shear modulus for $z<0$ and
ocean with zero shear modulus for $z>0$ (see Blaes et al.~1989 for the
Alfven wave case). This toy problem is of limited validity, since the 
density is not constant, and also since in reality the WKB assumption is violated (vertical wavelength in the ocean $\gg$ scale height),
but we can gain useful intuition without much effort. 
We consider a vertically propagating Hall wave incident
from below, and then include all possible reflected and transmitted waves.
The two circular polarization states are decoupled
in the boundary conditions, hence we need only include reflected Hall and
elastic-Alfven waves and a transmitted Alfven wave. We use the boundary 
conditions derived below to determine the amplitudes of reflected and 
transmitted waves. We leave this straightforward calculation as an exercise 
for the reader, and present the results. 
In the limit $B^2 \ll 4\pi \mu$ near the top of the crust, 
we find the amplitudes of the reflected Hall, reflected elastic-Alfven, and 
transmitted Alfven waves are
\be
\delta B_{\rm (ref\ Hall)} & \simeq & 
- \left( 1 + \frac{v_{\rm Hall}}{v_{\rm Alfven}} \right) \delta B_{\rm (inc\ Hall)}
\nonumber \\
\delta B_{\rm (ref\ Alfven)} & \simeq &
- \frac{v_{\rm Hall}v_{\rm Alfven}}{v_{\rm shear}^2} \delta B_{\rm (inc\ Hall)}
\nonumber \\
\delta B_{\rm (tran\ Alfven)} & \simeq &
- \frac{v_{\rm Hall}}{v_{\rm Alfven}} \delta B_{\rm (inc\ Hall)},
\ee
yielding an
energy transmission coefficient 
\be
T=2v_{\rm
Hall}/v_{\rm Alfven} & \simeq  & 2.4\times 10^{-12} \left[ B_{z,12} P({\rm yr})
\right]^{-1/2}
.
\label{eq:transmission}
\ee
Hence Hall
waves experience negligible damping due to wave leakage in this simple
case. Interestingly, we find that both the reflected and transmitted
Alfven waves have a very small amplitude compared to the Hall wave.
In this simple toy problem, the (observable) field in the liquid is
composed of the constant background field (generated in the core)
as well as the perturbation.  Since the amplitude of the perturbation
is infinitesimally small in the liquid as compared to the solid, the
observable external field may be somewhat different than the field in the
crust. If this behavior persists for more realistic field distributions,
the observable external field may not be the same as the field in the
crust; the neutron star is in effect ``hiding" its field. This effect is
of course larger when the currents are mainly in the crust rather than the
core.

If we relax the restriction of vertical propagation (i.e. allow
finite $k_\perp$), there is a qualitative change in the nature of the
waves. The low Hall wave frequency implies the reflected elastic-Alfven
wave becomes evanescent with e-folding distance $\sim k_\perp^{-1}$,
and the transmitted Alfven wave has wavevector nearly perpendicular
to the background magnetic field. Both polarization states must now
be included for the reflected elastic-Alfven wave and transmitted
Alfven wave. In addition, the two polarizations of reflected Hall wave
correspond to propagating and evanescent waves.  With so many waves
included, the analysis is complicated, and a nontrivial check of both
our numerical results and the boundary conditions derived below was that
energy was conserved over a large range in frequency and $k_\perp$. For
incident Hall wave of amplitude unity, the numerical results can
be summarized as : (i) the amplitude of the reflected propagating
Hall wave is $1 + {\cal O}\left(k_\perp/k_{\rm z,Hall}\right)$, (ii)
the amplitude of {\it all} other reflected waves scales as ${\cal
O}\left(k_\perp/k_{\rm z,Hall}\right)$, (iii) the transmission coeffient
is independent of $k_\perp$, and agrees in order of magnitude with
eq.\ref{eq:transmission}. Hence even in this more general case the crust
still acts as a nearly perfect resonant cavity, and the wave is essentially
unobservable since it's amplitude in the liquid is so small.

We now discuss the boundary conditions. We do not linearize the magnetic
field, but for simplicity we use the linearized shear stress. We use the
notation $[f]=(f)_{\rm solid}-(f)_{\rm liquid}$, and the normal to the 
boundary is in the $z$ direction. By integrating over a small region near the 
solid-liquid boundary we find 
\be
&& \left[B_z\right]  =  0 \hspace{2.0cm}\mbox{from $\divr \vec{B}=0$ }
\label{eq:Bz} \\
&& \left[J_z\right]  =  0 \hspace{2.0cm}\mbox{from $\divr \vec{J}=0$ } 
\label{eq:Jz}  \\
&& \left[ \xi_z \right] = 0  \hspace{2.0cm}\mbox{from finiteness of $\divr \vec{\xi}$ }
\label{eq:xizcont} \\
&& \left[(K-2\mu/3)\divr \vec{\xi} \delta_{iz} + \mu\left(\xi_{i,z}+\xi_{z,i}\right)
+ (4\pi)^{-1}\left(B_iB_z - \frac{1}{2} B^2 \delta_{iz}\right) \right]
+ g \xi_z \left[\rho\right] \delta_{iz}   =  0
\hspace{1.0cm} \mbox{from momentum equation} \label{eq:momentum} \\
&& \left[ -\frac{1}{c} \left\{ \left( \dot{\vec{\xi}}- \frac{\vec{J}}{n_e e} \right)
\times \vec{B} \right\}_\perp + \frac{\vec{J}_\perp}{\sigma} \right] = 0
\hspace{1.0cm}\mbox{from induction equation}. \label{eq:Eperp}
\ee
The last term in equation (\ref{eq:momentum}) arises from the buoyancy force
$-\vec{e}_r g \delta \rho = \vec{e}_r g \divr(\rho \xi_z)$, and 
we have ignored the perturbed gravitational potential.
Equation (\ref{eq:xizcont}) is required since $\divr \vec{\xi}$ would be infinite if
$\xi_z$ was discontinuous across an infinitesimally thin layer.

The Hall effect, like Ohmic diffusion, involves a higher spatial
derivative than the ion advection term in the induction equation.
Hence an additional boundary condition comes from the requirement that
these terms remain finite in equation (\ref{eq:Eperp}).  If the magnetic field
was discontinuous across the effectively infinitely thin boundary,
the Hall and Ohmic electric fields would diverge. Hence we must add the
boundary condition
\be
&& \left[ \vec{B}_\perp \right] = 0 \hspace{2.0cm}\mbox{from finite Hall voltage.}
\label{eq:Bperp}
\ee
This boundary condition is similar to the no-slip condition from viscous
hydrodynamics\footnote{At present, we have no boundary condition on 
$\vec{\xi}_\perp$. Including a viscous shear stress would require 
$\left[ \vec{\xi}_\perp \right] = 0$.}.

The horizontal stress boundary conditions can be greatly simplified to give
a useful result for the strain.
Taken together, equations (\ref{eq:Bz}) and (\ref{eq:Bperp}) imply $\left[ \vec{B} \right]
= 0$, which has the interesting consequence that the magnetic stress
drops out of the boundary condition in eq.\ref{eq:momentum}. Since $\mu=0$ in
the liquid, the horizontal components of the (symmetrized)
strain must vanish on the solid
side of the boundary,
\be
\left(\xi_{i,z}+\xi_{z,i}\right)_{\rm solid} & = & 0 \hspace{2.0cm}\mbox{for
$i=x,y$.}
\label{eq:strainbc}
\ee
As the magnetic field in the crust evolves by the Hall effect, equation (\ref{eq:strainbc})
is not satisfied in general over the entire crust. As the wave approaches
the boundary, it must adjust in order to satisfy this condition. On
qualitative grounds, we expect this would reflect the magnetic energy
even if the wave is not short lengthscale.  This result suggests that
even for the fully nonlinear case, {\it if} the currents are generated
in the crust there may be a significant difference between crustal field
and observable external field.

As discussed by Kingsep et al.~(1990), the higher spatial derivatives give
rise to a boundary layer effect in Hall-MHD (without the elastic shear
stress), with ion advection dominating on large scales and the Hall effect
on small scales (shorter than the ion penetration depth).  As a test
of the boundary conditions, we have redone the transmission-reflection
problem of Blaes et al.~(1989) for Alfen waves to escape from the crust,
now including the Hall term in the induction equation.  We find that
for typical Alfven wave frequencies we recover the Blaes et al.~(1989)
Alfven-wave transmission coefficient, consistent with the idea that
the Hall term only affects Alfven waves quite close to the boundary,
but cannot alter a global property such as transmission out of the crust.

Lastly, we note that if the currents are small in the liquid (say due to low
conductivity), or equivalently
if the field is near the vacuum solution, one finds an interesting boundary
condition on the field in the solid. Since $J_z=0$ in the liquid, and $[J_z]=0$,
$J_z \propto k_x B_y-k_y B_x = 0$ in the solid also. We can always choose 
$k_y=0$ without loss of generality, again implying $B_y=0$ in the solid at the boundary.

\end{document}